\documentclass[lettersize,journal]{IEEEtran}
\usepackage{amsmath,amsfonts}
\usepackage{amssymb}
\usepackage{algorithm}
\usepackage{algorithmic}
\usepackage{array}
\usepackage[caption=false,font=normalsize,labelfont=sf,textfont=sf]{subfig}
\usepackage{textcomp}
\usepackage{stfloats}
\usepackage{url}
\usepackage{verbatim}
\usepackage{graphicx}
\usepackage{cite}
\usepackage{multirow}
\usepackage{makecell}
\usepackage{color}
\usepackage{balance}

\begin{document}

\title{Mutual Coupling Aware Channel Estimation for \\ RIS-Aided Multi-User mmWave Systems}

\author{Tian~Qiu,~Ruidong~Li,~Cunhua~Pan,~\IEEEmembership{Senior~Member,~IEEE},~Taihao~Zhang,~Dongnan~Xia,~Changhong~Wang,~and~Hong~Ren,~\IEEEmembership{Member,~IEEE}

\thanks{Tian Qiu, Cunhua Pan, Taihao Zhang, Dongnan xia and Hong Ren are with National Mobile Communications Research Laboratory, Southeast University, Nanjing, China (e-mail:tianqiu, cpan, taihao, dnxia, hren@seu.edu.cn).}
\thanks{Ruidong Li is the co-first author. Ruidong Li and Changhong Wang are with Shandong Yunhai Guochuang Innovative Technology Co., Ltd., Jinan, China (e-mail:lird, wangchh01@inspur.com).}
\thanks{Corresponding author: Cunhua Pan.}}

\maketitle

\begin{abstract}
This paper proposes a three-stage uplink channel estimation protocol for reconfigurable intelligent surface (RIS)-aided multi-user (MU) millimeter-wave (mmWave) multiple-input single-output (MISO) systems, where both the base station (BS) and the RIS are equipped with uniform planar arrays (UPAs). The proposed approach explicitly accounts for the mutual coupling (MC) effect, modeled via scattering parameter multiport network theory. In Stage~I, a dimension-reduced subspace-based method is proposed to estimate the common angle-of-arrival (AoA) at the BS using the received signals across all users. In Stage~II, MC-aware cascaded channel estimation is performed for a typical user. The equivalent measurement vectors for each cascaded path are extracted and the one with the strongest energy (referred to as the reference column) is reconstructed using a compressed sensing (CS)-based approach. By leveraging the structure of the cascaded channel, the reference column is rearranged to estimate the AoA at the RIS, thereby reducing the computational complexity associated with estimating other columns. Additionally, the common angle-of-departure (AoD) at the RIS is also obtained in this stage, which reduces the pilot overhead for estimating the cascaded channels of other users in Stage~III. A Riemannian manifold optimization framework is proposed to design the RIS phase shift training matrix to improve performance against MC effect and outperform random phase scheme. Simulation results validate that the proposed method yields better performance than the MC-unaware and existing approaches in terms of estimation accuracy and pilot efficiency.
\end{abstract}

\section{Introduction}

\IEEEPARstart{R}{econfigurable} intelligent surface (RIS) has emerged as a promising technology for 6G-and-beyond wireless communications~\cite{ref1,ref2}. An RIS consists of a large number of low-cost passive reflective elements whose phase shifts can be tuned to enhance the received signal power and mitigate co-channel interference at the intended receiver~\cite{ref3,ref4}. Millimeter-wave (mmWave) multiple-input multiple-output (MIMO) systems~\cite{ref33,ref34} stand to benefit significantly from these advantages offered by RIS. However, fully exploiting the potential of RIS requires accurate channel state information (CSI), which remains challenging to obtain. The difficulty arises from the large number of antennas at the base station (BS) and reflective elements at the RIS, necessitating efficient pilot overhead reduction strategies. Furthermore, RIS elements are typically passive, lacking active radio frequency (RF) chains and signal processing capabilities, rendering conventional channel estimation techniques inapplicable~\cite{ref5,ref6}. More critically, emerging RIS hardware designs introduce additional impairments, most notably the mutual coupling (MC) effect between RIS elements, which significantly alters the cascaded channel characteristics~\cite{ref7,ref8,ref9,ref10}. Consequently, channel estimation for RIS-aided systems has become a focal point of intense research~\cite{ref29,ref30}.

Several studies have addressed channel estimation under sparse structured channel models, which exploit the inherent sparsity of high-frequency mmWave channels to reduce pilot overhead and improve estimation accuracy by leveraging compressed sensing (CS) techniques~\cite{ref11}, direction-of-arrival (DOA) estimation methods~\cite{ref12}, and sparse Bayesian learning (SBL) frameworks~\cite{ref13}. The authors of~\cite{ref14} proposed a CS-based channel estimation method by exploiting the inherent sparse structure of the user-RIS-BS cascaded channel. In~\cite{ref15}, the authors exploited the fact that the sparse channel matrices of the cascaded channels of all users share a common row-column-block sparsity structure arising from the shared RIS-BS channel, and proposed a multi-user (MU) joint iterative channel estimator based on this observation. Inspired by the common column-block sparsity property, the double-structured sparsity of the angular cascaded channels among users was exploited in~\cite{ref16} to propose a double-structured orthogonal matching pursuit (DS-OMP) algorithm, thereby reducing pilot overhead. A significant reduction in pilot overhead for MU mmWave systems was achieved in~\cite{ref17} by fully exploiting the correlation among different cascaded channels. The authors of~\cite{ref27} extended the channel strategy in~\cite{ref17} to the uniform planar array (UPA)-type MU-MIMO case.

Nevertheless, these works are based on the conventional RIS-aided channel model, which assumes that each RIS element radiates electromagnetic (EM) waves independently and neglects the nonlinear coupling between elements. Notable efforts have been dedicated to accurately modeling MC in RIS-aided communication systems. For instance, microwave multiport network theory extends fundamental circuit and network concepts to facilitate complex microwave analysis in modeling MC. A circuit-based communication model for RIS-aided wireless systems based on the mutual impedances between all radiating elements was introduced in~\cite{ref7}. The authors of~\cite{ref8} were the first to derive an MC-aware beyond-diagonal (BD) RIS-aided wireless communication model using scattering (\(S\)-parameter) and impedance (\(Z\)-parameter) representations, and proved their equivalence. The authors of~\cite{ref9} proposed new optimization algorithms based on the \(S\)-parameter multiport network model and discussed its advantages compared with its counterpart based on the \(Z\)-parameter representation. The authors of~\cite{ref10} employed scattering parameter network analysis to derive a physically and EM compliant yet straightforward and tractable RIS-aided communication model that fully accounts for the effects of impedance mismatching and mutual coupling at the transmitter, RIS, and receiver. Despite the importance of considering the MC effect in RIS-aided channel model, few studies have addressed channel estimation for RIS-aided systems in the presence of MC. The authors of~\cite{ref18} proposed a two-stage MC-aware channel estimator that performs effectively under strong MC conditions by employing CS and dictionary reduction (DR) techniques for an active RIS-aided multiple-input single-output (MISO) system. However, this work assumed only a single user, and is therefore unsuitable for practical multi-user scenarios as well as the ubiquitous connectivity envisioned in 6G systems.

Against the above background, we propose an MC-aware three-stage channel estimation protocol for RIS-aided MU-MISO mmWave communication systems with single-antenna users, where both the BS and the RIS are equipped with UPAs. This work advances RIS-aided MU-MISO channel estimation by explicitly incorporating the MC effect among RIS elements, which enables accurate characterization of cascaded channel under practical RIS deployment conditions. The main contributions of this paper are summarized as follows:
\begin{itemize}
    \item We propose an MC-aware three-stage channel estimation protocol for RIS-aided MU-MISO systems based on the \(S\)-parameter multiport network model. In Stage~I, the common angle-of-arrival (AoA) at the BS is estimated using the received signals from all users. In Stage~II, complete CSI estimation is performed for a typical user. After eliminating the common AoA influence, the reference column (i.e., the cascaded path with the strongest energy) is extracted and reformulated using an OMP-based approach. By exploiting the structure of the cascaded channel, the elements of the reference column are rearranged to estimate the AoA at the RIS of the typical user. This step reduces the computational complexity for estimating the other columns. Additionally, the common angle-of-departure (AoD) at the RIS is obtained, thereby significantly reducing the pilot overhead for estimating the cascaded channels of other users in Stage~III.
    \item In Stage~I of the protocol, we propose an effective dimension-reduced subspace-based method to estimate common AoA at the BS by processing the received signals from all users. The use of a UPA configuration in this paper necessitates extending subspace-based methods to two-dimensional (2-D) versions, which substantially increases computational complexity. To address this challenge, our proposed approach reduces computational burden while maintaining compatibility with existing one-dimensional (1-D) subspace-based algorithms by exploiting the UPA structure and leveraging the properties of the Kronecker product. Specifically, we evaluate the performance of root multiple signal classification (Root-MUSIC) and total least squares estimation of signal parameters via rotational invariance techniques (TLS-ESPRIT) algorithms for estimation in this stage.
    \item We optimize the equivalent RIS phase shift training matrices to enforce orthogonality among the columns of the equivalent dictionary and to enhance the performance of the OMP-based channel estimation, since MC inherently reduces the orthogonality of the equivalent dictionary. To this end, an efficient Riemannian manifold optimization framework is developed, which achieves performance gains compared to the random phase scheme. Furthermore, we analyze the pilot overhead and computational complexity of the proposed estimation protocol.
    \item We present a channel estimation scheme for the conventional RIS-aided cascaded channel model that neglects the MC effect. Applying this scheme to MC-affected signals provides a baseline for comparison with MC-aware methods, enabling a clear assessment of the impact of MC and the benefits of the proposed protocol.
\end{itemize}

The remainder of this paper is organized as follows. Section~\ref{model} introduces the system model. Section~\ref{MCestimation} details the proposed MC-aware three-stage channel estimation protocol. This section also presents the MC-unaware channel estimation scheme and an analysis of the pilot overhead and computational complexity. The design of phase shift training matrices is presented in Section~\ref{RIS}. Section~\ref{simulation} provides simulation results. Finally, Section~\ref{conclusion} concludes the paper.

\textit{Notations}: Boldface lowercase \(\mathbf{x}\) and uppercase \(\mathbf{X}\) denote vectors and matrices with \(\left[\mathbf{x}\right]_{m}\) and \(\left[\mathbf{X}\right]_{m,n}\) denoting the \(m\)-th and \((m,n)\)-th entry of \(\mathbf{x}\) and \(\mathbf{X}\), respectively. The \(m\)-th row and \(n\)-th column of matrix \(\mathbf{X}\) are represented by \(\mathbf{X}_{m,:}\) and \(\mathbf{X}_{:,n}\). For a matrix \(\mathbf{X}\) of arbitrary size, the symbols \(\mathbf{X}^*\), \(\mathbf{X}^\mathrm{T}\), \(\mathbf{X}^\mathrm{H}\), and \(\mathbf{X}^\dagger\) represent the conjugate, transpose, Hermitian, and pseudo-inverse of matrix \(\mathbf{X}\), respectively. \(\mathbf{X}^{-1}\) denotes the inverse of a square full-rank matrix \(\mathbf{X}\). The modulus of a scalar is denoted by \(|\cdot|\), the norm of a vector by \(\|\cdot\|\). The Euclidean norm of vector \(\mathbf{x}\) is denoted by \(\|\mathbf{x}\|_2\) and the Frobenius norm of matrix \(\mathbf{X}\) is denoted by \(\|\mathbf{X}\|_F\). \(\mathrm{Diag}(\mathbf{x})\) denotes a diagonal matrix with the entries of vector \(\mathbf{x}\) on its main diagonal. The vectorization operator \(\mathrm{vec}(\mathbf{X})\) stacks the columns of \(\mathbf{X}\) into a column vector, whereas \(\mathrm{mat}(\mathbf{X})\) reshapes \(\mathbf{X}\) into a matrix of specified dimensions following column-wise ordering. The expectation operator is denoted by \(\mathbb{E}\left[\cdot\right]\). Additionally, the Kronecker product, Hadamard product, Khatri-Rao product, and transposed Khatri-Rao product between two matrices \(\mathbf{X}\) and \(\mathbf{Y}\) are denoted by \(\mathbf{X} \otimes \mathbf{Y}\), \(\mathbf{X} \odot \mathbf{Y}\), \(\mathbf{X} \diamond \mathbf{Y}\) and \(\mathbf{X} \bullet \mathbf{Y}\), respectively. \(\lfloor x\rfloor\) rounds down to the nearest integer. 
\section{System Model}\label{model}

A narrow-band time-division duplex (TDD) mmWave system is considered, where \(K\) single-antenna users communicate with a BS equipped with an UPA comprising \(N = N_h \times N_v\) antennas, where \(N_h\) and \(N_v\) denote the numbers of horizontal and vertical elements. To improve communication efficiency, an RIS is deployed, which comprises a passive reflecting UPA with \(M = M_h \times M_v\) elements (\(M_h\) horizontal elements and \(M_v\) vertical elements). In this model, it is assumed that the direct channels between the users and the BS are blocked. In the following subsections, the RIS-BS and user–RIS subchannels are characterized and then the cascaded channel model is introduced while accounting for the MC effects among the elements of RIS.

\subsection{The RIS-BS Subchannel}

The narrow-band frequency-domain mmWave channel from the RIS to the BS, denoted as \(\mathbf{H} \in \mathbb{C}^{N\times M}\), is given by
\begin{align}
    \mathbf{H} = \sqrt{\frac{NM}{L}} \sum_{l=1}^{L} \tilde{\alpha}_{l} \mathbf{a}_{N} \left( \psi_{l}, \nu_{l} \right) \mathbf{a}_{M}^{\mathrm{H}} \left( \omega_{l}, \mu_{l} \right),
    \label{eq1}
\end{align}
where \(L\) denotes the number of propagation paths between the RIS and the BS, and the parameters \(\tilde{\alpha}_{l}\), \(\left( \psi_{l}, \nu_{l} \right)\), and \(\left( \omega_{l}, \mu_{l} \right)\) denote the complex channel gain, the AoA at the BS, and the AoD at the RIS for the \(l\)-th path in the RIS–BS channel, respectively. We further define the equivalent channel gain as \(\alpha_{l} \triangleq \tilde{\alpha}_{l}/\sqrt{L}\).

The Saleh–Valenzuela (SV) model is employed~\cite{ref19} to characterize the channel by exploiting the limited scattering characteristics of mmWave propagation. Define the array response vector (ARV) under a UPA configuration as \(\mathbf{a}_X(y,z) \in \mathbb{C}^{X\times1}\), where \(X = X_h\times X_v\), i.e.,
\begin{align}
    \mathbf{a}_X(z, y) = \mathbf{a}_{X_v}(z) \otimes \mathbf{a}_{X_h}(y),
    \label{eq2}
\end{align}
where \(\mathbf{a}_{X_v}(z) = 1/\sqrt{X_v} \left[ 1, e^{-i2\pi z}, \cdots, e^{-i2\pi(X_v-1)z} \right]^\mathrm{T}\) and \(\mathbf{a}_{X_h}(y) = 1/\sqrt{X_h} \left[ 1, e^{-i2\pi y}, \cdots, e^{-i2\pi(X_h-1)y} \right]^\mathrm{T}\), denoting the steering vectors for the vertical (\(z\)-axis) and horizontal (\(y\)-axis) UPA dimensions, respectively. The variables \(z\) and \(y\) represent the equivalent spatial frequencies in the vertical and horizontal planes of the UPA, respectively.

Define the elevation angle \(\varrho \in \left[-90^\circ, 90^\circ \right)\) and the azimuth angle \(\varsigma \in \left[ -180^\circ,180^\circ \right)\) of the incoming signal. The spatial frequency pair \((z,y)\) is related to the physical angle pair \((\varrho,\varsigma)\) as follows, \(z = d \sin (\varrho)/\lambda_c \) and \( y = d \sin (\varsigma)\cos(\varrho)/\lambda_c \), where \(\lambda_c=c/f_c\) is the carrier wavelength and \(d\) is the inter-element spacing of the UPA. If \(d \le \lambda_c/2\), a one-to-one mapping exists between spatial frequencies and physical angles on one side of the UPA~\cite{ref17}. We assume this relationship holds throughout the paper, and refer to the steering vector arguments interchangeably as either physical angles or spatial frequencies.

Moreover, the RIS-BS channel in Eq.~\eqref{eq1} can be written in a more compact way as
\begin{align}
    \mathbf{H} = \mathbf{A}_{N} \mathbf{\Lambda} \mathbf{A}_{M}^{\mathrm{H}},
    \label{eq3}
\end{align}
where
\begin{subequations}
    \begin{align}
        \mathbf{A}_{N} & = \left[ \mathbf{a}_{N} \left( \psi_{1}, \nu_{1} \right), \cdots, \mathbf{a}_{N} \left( \psi_{L}, \nu_{L} \right) \right] \in \mathbb{C}^{N\times L},
        \label{eq4a} \\
        \mathbf{\Lambda} & = \mathrm{Diag} \left( \alpha_{1}, \cdots, \alpha_{L} \right) \in \mathbb{C}^{L\times L},
        \label{eq4b} \\
        \mathbf{A}_{M} & = \left[ \mathbf{a}_{M} \left( \omega_{1}, \mu_{1} \right), \cdots, \mathbf{a}_{M} \left( \omega_{L}, \mu_{L} \right) \right] \in \mathbb{C}^{M\times L}.
        \label{eq4c}
    \end{align}
\end{subequations}
    
\subsection{The User-RIS Subchannel}

Similarly, the frequency-domain channel from user \(k\) to the RIS, denoted as \(\mathbf{h}_{k} \in \mathbb{C}^{M \times1}\), can be represented as
\begin{align}
    \mathbf{h}_{k} = \sqrt{\frac{M}{J_k}} \sum_{j=1}^{J_{k}} \tilde{\beta}_{k,j} \mathbf{a}_{M} \left( \varphi_{k,j}, \theta_{k,j} \right), \forall k \in \mathcal{K},
    \label{eq5}
\end{align}
where \(J_k\) denotes the number of propagation paths between user \(k\) and the RIS, and \(\beta_{k,j}\) is the complex channel gain of the \(j\)-th path in the user–RIS channel. The pair \(\left( \varphi_{k,j}, \theta_{k,j} \right)\) specifies the AoA of the \(j\)-th path from user \(k\) to the RIS. The set of users is defined as \(\mathcal{K} = \{1, 2, \cdots, K\}\). Similarly, we define the equivalent channel gain as \(\beta_{k,j} \triangleq \tilde{\beta}_{k,j}/\sqrt{J_k}\). The \(k\)-th user-RIS channel in Eq.~\eqref{eq5} can also be represented in a more compact form as
\begin{align}
    \mathbf{h}_{k} = \mathbf{A}_{M,k} \boldsymbol{\beta}_{k},
    \label{eq6}
\end{align}
where
\begin{subequations}
    \begin{align}
        \mathbf{A}_{M,k} & = \left[ \mathbf{a}_{M} \left( \varphi_{k,1},  \theta_{k,1} \right), \cdots, \mathbf{a}_{M} \left( \varphi_{k,J_{k}}, \theta_{k,J} \right) \right] \in \mathbb{C}^{M\times J_{k}},
        \label{eq7a} \\
        \boldsymbol{\beta}_{k} & = \left[ \beta_{k,1}, \cdots, \beta_{k,J_k} \right]^\mathrm{T} \in \mathbb{C}^{J_k\times 1}.
        \label{eq7b}
    \end{align}
\end{subequations}

\subsection{Cascaded Channel Model}

Since direct channels between the users and BS are assumed to be blocked, hence, the conventional RIS-aided channel model without MC between elements of the RIS is given by
\begin{align}
    \mathbf{h}_{\mathrm{cv},k} = \mathbf{H} \mathbf{\Gamma} \mathbf{h}_{k},
    \label{eq8}
\end{align}
where \( \mathbf{\Gamma} = \mathrm{Diag}(\boldsymbol{\gamma})\), with \(\boldsymbol{\gamma} \in \mathbb{C}^{M\times 1}\) denoting the RIS phase shift vector. Eq.~\eqref{eq8} reveals that the received signal at the BS is a linear combination of reflections from each element. This model assumes that each RIS element reflects incident EM waves independently, neglecting any MC effects. When strong MC occurs among the elements of RIS, the assumption can cause model mismatch and hinder tasks such as channel estimation, thereby degrading the overall communication performance. Based on the \(S\)-parameter multiport network theory, an MC–aware communication model has been recently derived and validated~\cite{ref8,ref9,ref18}. Incorporating MC into the RIS response, the channel model in Eq.~\eqref{eq8} is reformulated as:
\begin{align}
    \mathbf{h}_{\mathrm{mc},k} = \mathbf{H} \left( \mathbf{\Gamma}^{-1}-\mathbf{S} \right)^{-1} \mathbf{h}_{k},
    \label{eq9}
\end{align}
where \(\mathbf{S} \in \mathbb{C}^{M\times M}\) denotes the RIS scattering matrix capturing MC among the elements of RIS. The scattering and impedance matrices are related by \(\mathbf{S} = \left( \mathbf{Z} + Z_{0} \mathbf{I}_M \right)^{-1} \left( \mathbf{Z} - Z_{0} \mathbf{I}_M \right)\)~\cite{ref8,ref9}, where \(\mathbf{Z} \in \mathbb{C}^{M\times M}\) is the RIS impedance matrix and \(Z_0\) is the characteristic impedance~\cite{ref7,ref28} \footnote{Unless otherwise specified, channel parameters implicitly include MC and will be denoted without the subscript \(\mathrm{mc}\).}. Specifically, we consider two arbitrary RIS radiating elements \(\chi=\{p,q\}\), each characterized by a length-radius pair \((l_\chi,a_\chi)\) and spatial location \(\mathbf{r}_\chi = x_\chi \hat{\mathbf{x}} + y_\chi \hat{\mathbf{y}} + z_\chi \hat{\mathbf{z}}\). The impedance matrix is constructed as \(\mathbf{Z} = \left[ Z_{qp} \right]_{M\times M}\), where \(Z_{qp}\) represents the MC effect of element \(p\) on element \(q\). The explicit expression of \(Z_{qp}\) is provided in Eq.~\eqref{Z}, where the normalized current on the surface of \(\chi\) is modeled as \({I}_{z,\chi}(z) = {\sin\left(k_0\left({l_{\chi}}/{2} - |z-z_{\chi}|\right)\right)} /{\sin\left({k_0 l_{\chi}}/{2}\right)}\).

\begin{figure*}[t] 
\begin{align}
    Z_{qp} = \frac{j\eta_0}{4\pi k_0} \int_{z_q-l_q/2}^{z_q+l_q/2} \int_{z_p-l_p/2}^{z_p+l_p/2}
    {I}_{z,p}(z_1) {I}_{z,q}(z_2) \frac{ e^{-j k_0 R(z_1,z_2)}}{R(z_1,z_2)}
    & \left(
    k_0^2 - \frac{j k_0}{R(z_1,z_2)} - \frac{ k_0^2 (z_2 - z_1)^2 + 1 }{ R^2(z_1,z_2) }
    \right. \notag \\
    & \left. + \frac{ 3 j k_0 (z_2 - z_1)^2 }{ R^3(z_1,z_2) } + \frac{ 3 (z_2 - z_1)^2 }{ R^4(z_1,z_2) }
    \right)
    \mathrm{d}z_1 \mathrm{d}z_2.
\label{Z}
\end{align}
\hrule
\end{figure*}
Here, \(\eta_0=\sqrt{\mu_0/\epsilon_0}\) denotes the intrinsic impedance of free space, and
\(k_0=2\pi/\lambda_c\) is the wavenumber, where \(\mu_0\) and \(\epsilon_0\) are the magnetic permeability and the electric permittivity, respectively. The function \(R(z_1,z_2)\) represents the distance between \(\mathbf{r}_p\) and \(\mathbf{r}_q\), and is defined as

\begin{align}
& R(z_1,z_2) = \sqrt{\rho^2 + (z_2 - z_1)^2}, \notag \\
&   \begin{cases}
        \rho = a_p,  & \text{if } p = q, \\
        \rho = \sqrt{(x_q - x_p)^2 + (y_q - y_p)^2}, & \text{if } p \neq q.
        \end{cases}
\label{R}
\end{align}
Comparing Eq.~\eqref{eq8} with Eq.~\eqref{eq9}, it can be observed that, once the MC effect is taken into account, RIS elements no longer radiate EM waves independently and instead exhibit nonlinear coupling among elements. Specifically, the RIS response matrix loses its diagonal structure under MC, and the reflections from different RIS elements become strongly coupled. This coupling fundamentally alters both the structure and dimensionality of the cascaded user-RIS-BS channel, invalidates the diagonal property exploited by conventional RIS channel estimation methods, and significantly degrades the orthogonality of the equivalent sensing matrix. As a result, standard OMP-based approaches developed for MC-unaware models are no longer directly applicable. As demonstrated in the subsequent analysis, these effects introduce additional challenges for reliable channel estimation.

\subsection{The Recived Siganl}

Denote \(s_k(t)\) as the pilot signal of the \(k\)-th user. To eliminate inter-user interference, the users employ a one-by-one transmission scheme. Accordingly, during the uplink transmission, the received signal from user \( k \) in time slot \( t \), where \( 1 \leq t \leq \tau_k \), can be expressed as
\begin{align}
    \mathbf{y}_k(t) 
    = & \mathbf{H} \left( \mathbf{\Gamma}_{t}^{-1} - \mathbf{S} \right)^{-1}  \left( \sqrt{p} \mathbf{h}_{k}  s_k(t) + \mathbf{n}_{2,k}(t) \right) + \mathbf{n}_{1,k}(t) \notag \\
    \triangleq & \sqrt{p} \mathbf{H} \left( \mathbf{\Gamma}_{t}^{-1} - \mathbf{S} \right)^{-1} \mathbf{h}_{k} s_k(t) + \mathbf{n}_{k}(t),
    \label{eq10}
\end{align}
where \(\mathbf{n}_{k}(t) = \mathbf{H} \left( \mathbf{\Gamma}_{t}^{-1} - \mathbf{S} \right)^{-1} \mathbf{n}_{2,k}(t) + \mathbf{n}_{1,k}(t)\) and \(p\) represents transmit power of each user. \(\mathbf{n}_{1,k}(t) \in \mathbb{C}^{N\times 1}\) and \(\mathbf{n}_{2,k}(t) \in \mathbb{C}^{M\times 1}\) model the thermal noise at the BS and RIS, respectively, which are distributed as \(\mathbf{n}_{1,k}(t) \sim \mathcal{CN} \left( 0,\sigma_1^2 \mathbf{I}_N \right)\) and \(\mathbf{n}_{2,k}(t) \sim \mathcal{CN} \left( 0, \sigma_2^2 \mathbf{I}_M \right)\). In most existing works, \(\mathbf{n}_{2,k}(t)\) is neglected because its contribution is comparatively small relative to the noise power at the BS after reflection and going through backward channel. However, due to MC, the noise introduced at the RIS cannot be ignored. Furthermore, incorporating this component enables a more accurate quantification of signal-to-noise ratio (SNR) degradation in RIS-aided systems by capturing correlated noise effects at the RIS.

Assuming that the pilot symbols satisfy \(s_k(t) = 1\) for \(1  \le t \le \tau_k\) and defining \(\mathbf{B}_{t} \triangleq \left( \mathbf{\Gamma}_t^{-1} - \mathbf{S} \right)^{-1}\), the received signal can be rewritten as
\begin{align}
    & \mathbf{y}_k(t) = \sqrt{p} \mathbf{H} \mathbf{B}_{t} \mathbf{h}_{k} + \mathbf{n}_{k}(t) \notag \\
    & = \sqrt{p} \mathrm{vec} \left( \mathbf{A}_{N} \mathbf{\Lambda} \mathbf{A}_{M}^{\mathrm{H}} \mathbf{B}_{t} \mathbf{A}_{M,k} \boldsymbol{\beta}_{k} \right) + \mathbf{n}_{k}(t) \notag \\
    & = \sqrt{p} \left( \boldsymbol{\beta}_{k}^{\mathrm{T}} \otimes \mathbf{A}_{N} \mathbf{\Lambda} \right) \mathrm{vec} \left(\mathbf{A}_{M}^{\mathrm{H}} \mathbf{B}_{t} \mathbf{A}_{M,k} \right) + \mathbf{n}_{k}(t) \notag \\
    & = \sqrt{p} \left( 1 \otimes \mathbf{A}_{N} \right) \left( \boldsymbol{\beta}_{k}^{\mathrm{T}} \otimes \mathbf{\Lambda} \right) \left( \mathbf{A}_{M,k}^{\mathrm{T}} \otimes \mathbf{A}_{M}^{\mathrm{H}} \right) \mathrm{vec} \left( \mathbf{B}_{t} \right) + \mathbf{n}_{k}(t) \notag \\
    & \triangleq \sqrt{p} \mathbf{A}_{N} \left( \boldsymbol{\beta}_{k}^{\mathrm{T}} \otimes \mathbf{\Lambda} \right) \left( \mathbf{A}_{M,k}^{\mathrm{T}} \otimes \mathbf{A}_{M}^{\mathrm{H}} \right) \boldsymbol{\eta}_{t} + \mathbf{n}_{k}(t),
    \label{eq11}
\end{align}
where \(\boldsymbol{\eta}_t = \mathrm{vec} \left( \mathbf{B}_{t} \right)\) is the RIS phase shift vector incorporating the MC effect. Stacking the \(\tau_k\) received signal vectors across time slots yields the received signal matrix \(\mathbf{Y}_{k} \in \mathbb{C}^{N \times \tau_{k}}\), which can be written as
\begin{align}
    \mathbf{Y}_{k} = \left[ \mathbf{y}_{k}(1), \cdots, \mathbf{y}_{k} \left( \tau_{k} \right) \right] \triangleq \sqrt{p} \mathbf{G}_{k} \mathbf{\Theta}_{k} + {\mathbf{N}}_{k},
    \label{eq12}
\end{align}
where \( \mathbf{G}_{k} = \mathbf{A}_{N} \left( \boldsymbol{\beta}_{k}^{\mathrm{T}} \otimes \mathbf{\Lambda} \right) \left( \mathbf{A}_{M,k}^{\mathrm{T}} \otimes \mathbf{A}_{M}^{\mathrm{H}} \right)\) defines the equivalent MC-aware cascaded user–RIS–BS channel for user \(k\), which we aim to estimate. Moreover, \(\mathbf{\Theta}_{k} = \left[ \boldsymbol{\eta}_{1}, \cdots, \boldsymbol{\eta}_{\tau_{k}} \right] \in \mathbb{C}^{M^2 \times \tau_{k}}\) denotes the MC-aware phase shift training matrix of the RIS and \(\mathbf{N}_{k} = \left[ \mathbf{n}_{k}(1), \cdots, \mathbf{n}_{k} \left( \tau_{k} \right) \right] \in \mathbb{C}^{N \times \tau_{k}}\) represents the noise matrix for user \(k\). The high dimensionality of \(\mathbf{\Theta}_{k}\), together with the consideration of MC, will degrade the orthogonality of the equivalent dictionary.

\section{MC-aware Channel Estimation}\label{MCestimation}

\subsection{Channel Estimation Protocol}

\begin{figure}[H]
    \centering
    \includegraphics[width=3.5in]{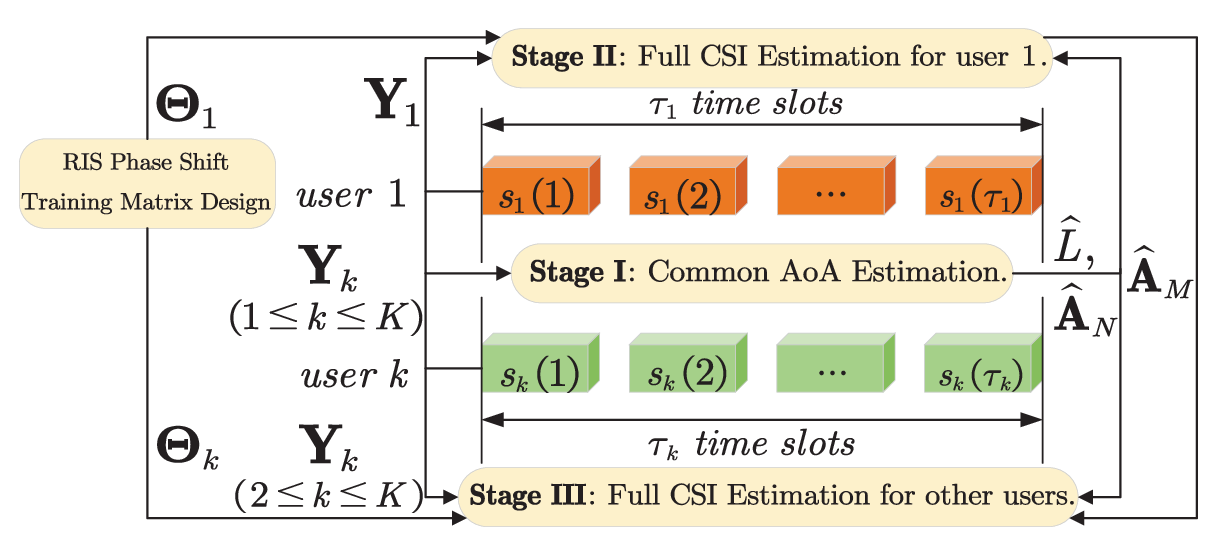}
    \caption{The proposed MC-aware three-stage channel estimation protocol.}
\label{protocol}
\end{figure}

Fig.~\ref{protocol} illustrates the proposed MC-aware three-stage uplink channel estimation protocol. As discussed above and further elaborated in Section~\ref{unaware}, the MC effect disrupts the diagonal structure of the RIS response in conventional RIS-aided cascaded channel models. This results in higher-dimensional equivalent dictionary matrices for channel estimation~\cite{ref18}, which in turn leads to increased pilot overhead to achieve satisfactory estimation accuracy. To address this challenge, the proposed protocol adopts a staged estimation strategy that exploits cascaded channel sparsity and the common RIS-BS channel shared by all users, thereby significantly reducing pilot overhead while maintaining high estimation accuracy.

\textit{Stage~I: Common AoA Estimation at the BS.}  
In Stage~I, a dimension-reduced subspace-based method is employed to estimate the common AoA at the BS. By exploiting the UPA structure and the Kronecker product property, the 2-D AoA estimation problem is decomposed into two independent 1-D estimations along the horizontal and vertical dimensions. Moreover, the common AoA is estimated using the received signals from all users, which improves robustness and mitigates error propagation in subsequent stages.
    
\textit{Stage~II: MC-Aware Cascaded Channel Estimation for the Typical User.}  
In Stage~II, MC-aware cascaded channel estimation is performed for user~1, selected as the user closest to the RIS due to its lower path loss and stronger received signal at the BS. After eliminating the common AoA components by exploiting the asymptotic orthogonality of steering matrix in large-scale arrays, equivalent measurement vectors corresponding to individual cascaded paths are extracted. The reference column (i.e., the path with the strongest energy) is first reconstructed using an OMP-based method. To further enhance estimation accuracy, an effective Riemannian manifold optimization framework is proposed to design the RIS phase shift training matrix. By leveraging the Kronecker product structure of the cascaded channel, the reference column is rearranged to estimate the AoA at the RIS for user~1 according to the index order, which reduces the dimension of the equivalent dictionary matrix and the computational complexity for estimating the other cascaded paths. In addition, the common AoD at the RIS is obtained in this stage.

\textit{Stage~III: MC-Aware Cascaded Channel Estimation for Other Users.}  
Finally, in Stage~III, the estimates obtained from the previous two stages are utilized to reconstruct the structural characteristics of the common RIS-BS channel shared by all users. This enables efficient MC-aware cascaded channel estimation for the other users with substantially reduced pilot overhead and computational complexity. The following subsections present a detailed elaboration of the proposed MC-aware channel estimation protocol.

\subsection{Stages~I and~II: Channel Estimation for Typical User}

This subsection outlines a comprehensive procedure for estimating the full CSI of the typical user, denoted as user~1. In Stage~I, a dimension-reduced subspace-based method is employed to estimate the common AoA at the BS by processing the received signal of all users, thereby mitigating error propagation. In Stage~II, we estimate the MC-aware cascaded channel for user~1 by reformulating the equivalent measurement vector for each cascaded path. The channel is then reconstructed using an OMP-based approach. This stage also yields the estimate of the common AoD at the RIS, which subsequently reduces the pilot overhead for the other users.

\subsubsection{Estimation of Common AoA at the BS}\label{AoABS}

From Eq.~\eqref{eq12}, the received signal matrix \(\mathbf{Y}_k \in \mathbb{C}^{N\times\tau_k}\) can be expressed as
\begin{align}
    \mathbf{Y}_{k} = \sqrt{p} \mathbf{G}_{k} \mathbf{\Theta}_{k} + {\mathbf{N}}_{k} \triangleq \sqrt{p} \mathbf{A}_{N} \tilde{\mathbf{S}}_{k} + \mathbf{N}_{k},
\label{eq13}
\end{align}
where \(\tilde{\mathbf{S}}_{k} = \left( \boldsymbol{\beta}_{k}^\mathrm{T} \otimes \mathbf{\Lambda} \right)\left( \mathbf{A}_{M,k}^\mathrm{T} \otimes \mathbf{A}_{M}^{\mathrm{H}} \right) \mathbf{\Theta}_{k} \in \mathbb{C}^{L\times\tau_k}\) represents the equivalent transmitted signal matrix for user~\(k\). Consequently, subspace-based methods exhibit superior performance under this standard signal models. However, the UPA configuration in this paper necessitates their extension to the 2-D version. Such an extension substantially increases the algorithm’s computational complexity. For instance, the 2-D Root-MUSIC algorithm requires evaluating determinants whose dimensions scale with the array size~\cite{ref20}. To address this, a dimension-reduced subspace-based method is proposed to reduce computational burden while remaining compatible with existing 1-D estimation algorithms.

Exploiting the structure of the UPA and leveraging the properties of the Kronecker product, we derive a dimension-reduced received signal matrix \(\mathbf{Y}_{k,h} \in \mathbb{C}^{N_h \times \tau_k N_v}\) for the horizontal dimension to estimate the horizontal angles \(\{\nu_l\}_{l=1}^{L}\), which is achieved via row extraction as follows
\begin{align}
    & \left[ \mathbf{Y}_{k,h} \right]_{:,1+(v-1)\tau_k:v\tau_k} 
    = \left[ \mathbf{Y}_{k} \right]_{1+(v-1)N_h:vN_h,:} \notag \\
    & = \sqrt{p} \left[ \mathbf{A}_N \right]_{1+(v-1)N_h:vN_h,:} \tilde{\mathbf{S}}_{k} + \left[\mathbf{N}_{k}\right]_{1+(v-1)N_h:vN_h,:} \notag \\
    & \triangleq \sqrt{p} \left[ \zeta_h \left(v, 1\right) \mathbf{a}_{N_h}(\nu_1), \cdots, \zeta_h \left(v, L\right) \mathbf{a}_{N_h}(\nu_L) \right] \tilde{\mathbf{S}}_{k} + \mathbf{N}_{k,h}^v \notag \\
    & = \sqrt{p} \mathbf{A}_{N_h} \mathrm{Diag} \left( \zeta_h \left(v, 1\right), \cdots, \zeta_h \left(v, L\right) \right) \tilde{\mathbf{S}}_{k} + \mathbf{N}_{k,h}^v,
    \label{eq14}
\end{align}
where \(\zeta_h \left(v, l\right) = e^{-i 2\pi (v-1)\psi_l}\) with \(v = 1, 2, \cdots, N_v\) denoting the row index, \(\mathbf{A}_{N_h} \triangleq \left[ \mathbf{a}_{N_h}(\nu_1), \cdots, \mathbf{a}_{N_h}(\nu_L) \right]\) denotes the horizontal AoA steering matrix, and \(\mathbf{N}_{k,h} = \left[ \mathbf{N}_{k,h}^1, \cdots, \mathbf{N}_{k,h}^{N_v} \right]\) represents the corresponding noise. Defining \(\tilde{\mathbf{S}}_{k,h}^v = \mathrm{Diag} \left( \zeta_h \left(v, 1\right), \cdots, \zeta_h \left(v, L\right) \right) \tilde{\mathbf{S}}_{k}\) yields \(\tilde{\mathbf{S}}_{k,h} = \left[\tilde{\mathbf{S}}_{k,h}^1, \cdots, \tilde{\mathbf{S}}_{k,h}^{N_v} \right] \in \mathbb{C}^{L\times N_v \tau_{k}}\). Thus, the 1-D standard signal model is given by \(\mathbf{Y}_{k,h} = \sqrt{p} \mathbf{A}_{N_h} \tilde{\mathbf{S}}_{k,h} + \mathbf{N}_{k,h}\). To enhance subspace-based methods by eliminating error propagation and increasing the equivalent received signal snapshots, we concatenate the dimension-reduced received matrices of all users into \(\mathbf{Y}_{h} = \left[ \mathbf{Y}_{1, h}, \cdots, \mathbf{Y}_{K, h} \right] \in \mathbb{C}^{N_h\times N_v \sum_{k=1}^{K} \tau_{k}}\), which is then used to compute the received signal covariance matrix
\begin{align}
    \mathbf{R}_{h} 
    & = \mathbb{E} \left[ \mathbf{Y}_{h} \mathbf{Y}_{h}^{\mathrm{H}} \right] \notag \\
    & = p \mathbf{A}_{N_{h}} \mathbb{E} \left[ \tilde{\mathbf{S}}_{h} \tilde{\mathbf{S}}_{h}^{\mathrm{H}} \right] \mathbf{A}_{N_{h}}^{\mathrm{H}} + \mathbb{E} \left[ \mathbf{N}_{h} \mathbf{N}_{h}^{\mathrm{H}} \right] ,
    \label{eq15}
\end{align}
where \(\tilde{\mathbf{S}}_{h} = \left[ \tilde{\mathbf{S}}_{1,h}, \cdots, \tilde{\mathbf{S}}_{K,h} \right] \in \mathbb{C}^{L\times N_v \sum_{k=1}^{K} \tau_{k}}\) and \(\mathbf{N}_{h} = \left[ \mathbf{N}_{1, h}, \cdots, \mathbf{N}_{K, h} \right] \in \mathbb{C}^{N_h\times N_v \sum_{k=1}^{K} \tau_{k}}\) represents the equivalent transmitted signal and noise across all users, respectively. 

Since the actual received signal matrix has finite length, eigenvalue decomposition is typically performed using the maximum likelihood estimation (MLE) of the signal covariance matrix, given by
\begin{align}
    \mathbf{R}_{h} = \frac{1}{N_v \sum_{k=1}^{K} \tau_{k}} \mathbf{Y}_{h} \mathbf{Y}_{h}^{\mathrm{H}}.
    \label{eq16}
\end{align}
By performing eigenvalue decomposition of \(\mathbf{R}_{h}\), we identify the signal subspace spanned by the eigenvectors associated with the \(L\) largest eigenvalues. The value \(L\) also corresponds to the number of common AoAs, which equals the number of propagation paths between the RIS and the BS\footnote{While the eigenvalues corresponding to the signal subspace are typically much larger than those of the noise subspace, information-theoretic criteria such as the minimum description length (MDL) and Akaike information criterion (AIC) can be employed to more accurately determine the number of signal sources.}. The same procedure is used to estimate the steering vectors for the vertical (\(z\)-axis) UPA dimension. It enables estimation of the common AoA steering matrix \(\widehat{\mathbf{A}}_{N} = \widehat{\mathbf{A}}_{N_v} \diamond \widehat{\mathbf{A}}_{N_h}\) shared by all users. Numerous studies have demonstrated the effectiveness of subspace-based algorithms~\cite{ref31}, which can achieve high estimation accuracy with acceptable computational complexity when compared with classical discrete Fourier transform (DFT)-based approaches~\cite{ref32}. In this work, two representative subspace-based methods, Root-MUSIC and TLS-ESPRIT, are adopted for common AoA estimation. Algorithm~\ref{alg1} outlines the estimation procedure, while the detailed subspace operations follow the standard formulations in~\cite{ref21,ref22}.

\begin{algorithm}
\caption{Dimension-reduced Subspace-Based Common AoA Estimation}
\label{alg1}
\begin{algorithmic}[1]  
\REQUIRE \(\mathbf{Y}_{k}\), \(1 \leq k \leq K\).
\STATE Calculate the horizontal dimension-reduced signal matrix \(\mathbf{Y}_{k, h}\) from \(\mathbf{Y}_{k}\) (\(1 \leq k \leq K\)) according to Eq.~\eqref{eq14}.
\STATE Concatenate the dimension-reduced covariance matrices across all users to obtain \(\mathbf{Y}_{h}= \left[ \mathbf{Y}_{1, h}, \cdots, \mathbf{Y}_{K, h} \right]\).
\STATE Estimate \(\mathbf{R}_{h}\) of \(\mathbf{Y}_{h}\) according to Eq.~\eqref{eq16}.
\STATE Perform the eigenvalue decomposition of \(\mathbf{R}_{h}\)
\begin{align}
    \mathbf{R}_{h} = \mathbf{U}_{S, h} \mathbf{\Sigma}_{S, h} \mathbf{U}_{S, h}^{\mathrm{H}} + \mathbf{U}_{N, h} \mathbf{\Sigma}_{N, h} \mathbf{U}_{N}^{\mathrm{H}},
    \label{eq17}
\end{align}
where \(\mathbf{U}_{S, h} \mathbf{\Sigma}_{S, h} \mathbf{U}_{S, h}^{\mathrm{H}}\) represents the signal portion and \(\mathbf{U}_{N, h} \mathbf{\Sigma}_{N, h} \mathbf{U}_{N, h}^{\mathrm{H}}\) represents the noise portion.
\STATE Sort the eigenvalues and determine the number of propagation paths in the RIS-BS channel, denoted as \(\widehat{L}\).
\STATE Apply the subspace-based algorithm to estimate the common AoAs in the horizontal dimension \(\{ \widehat{\nu}_{l} \}_{l=1}^{\widehat{L}}\) and construct the corresponding steering matrix \(\widehat{\mathbf{A}}_{N_h} = \left[ \mathbf{a}_{N_h}(\widehat{\nu}_{1}), \cdots, \mathbf{a}_{N_h}(\widehat{\nu}_{L}) \right]\).
\STATE Repeat the similar procedure for the vertical dimension to estimate \(\{\widehat{\psi}_{l}\}_{l=1}^{\widehat{L}}\) and construct the steering matrix \(\widehat{\mathbf{A}}_{N_v} = \left[ \mathbf{a}_{N_v}(\widehat{\psi}_{1}), \cdots, \mathbf{a}_{N_v}(\widehat{\psi}_{L}) \right]\).
\ENSURE \(\widehat{\mathbf{A}}_{N} = \widehat{\mathbf{A}}_{N_v} \diamond \widehat{\mathbf{A}}_{N_h}\).
\end{algorithmic}
\end{algorithm}

\subsubsection{Estimation of Common AoD at the RIS}\label{AoDRIS}

Let \( \Delta \mathbf{A}_{N} \triangleq \widehat{\mathbf{A}}_{N} - \mathbf{A}_{N} \) denotes the estimation error associated with the common AoA estimates. Exploiting the asymptotic orthogonality of the steering matrix for large-scale arrays, as established in~\cite{ref17}, we use the identity \( \widehat{\mathbf{A}}_{N}^\mathrm{H} \mathbf{A}_{N} = N \mathbf{I}_{L} + \left( \Delta \mathbf{A}_{N} \right)^\mathrm{H} \mathbf{A}_{N}\). Thus, the linear transformation \( \frac{1}{N \sqrt{p}} \widehat{\mathbf{A}}_{N}^{\mathrm{H}} \) can be applied to the received signal \(\mathbf{Y}_k\) to :
\begin{align}
    \frac{1}{N \sqrt{p}} \widehat{ \mathbf{A}}_{N}^{\mathrm{H}} \mathbf{Y}_{k}
    = \left( \boldsymbol{\beta}_{k}^\mathrm{T} \otimes \mathbf{\Lambda} \right)\left( \mathbf{A}_{M,k}^\mathrm{T} \otimes \mathbf{A}_{M}^{\mathrm{H}} \right) \mathbf{\Theta}_{k} + \breve{\mathbf{N}}_{k}^{\mathrm{H}} ,
    \label{eq18}
\end{align}
where \( \breve{\mathbf{N}}_{k}^{\mathrm{H}} \triangleq \frac{1}{ N \sqrt{p}} \widehat{\mathbf{A}}_{N}^{\mathrm{H}} \mathbf{N}_{k} + \frac{1}{N} \left( \Delta \mathbf{A}_{N} \right)^\mathrm{H} \mathbf{A}_{N} \left( \boldsymbol{\beta}_{k}^\mathrm{T} \otimes \mathbf{\Lambda} \right) \) \(\left( \mathbf{A}_{M,k}^\mathrm{T} \otimes \mathbf{A}_{M}^{\mathrm{H}} \right) \mathbf{\Theta}_{k}\) denotes the corresponding noise and the second term characterizes the residual error propagation introduced by common AoA estimates. Now, we define the transpose of \( \frac{1}{N \sqrt{p}} \widehat{\mathbf{A}}_{N}^{\mathrm{H}} \mathbf{Y}_{k} \) as the equivalent measurement matrix \( \breve{\mathbf{Y}}_{k} \in \mathbb{C}^{\tau_{k} \times L} \) for user \(k\), given by
\begin{align}
    \breve{\mathbf{Y}}_{k}  
    & \triangleq \left( \frac{1}{N \sqrt{p}} \widehat{\mathbf{A}}_{N}^{\mathrm{H}} \mathbf{Y}_{k} \right)^{\mathrm{H}} \notag \\
    & = \mathbf{\Theta}_{k}^{\mathrm{H}} \left( \mathbf{A}_{M,k}^\mathrm{*} \otimes \mathbf{A}_{M} \right) \left( \boldsymbol{\beta}_{k}^\mathrm{*} \otimes \mathbf{\Lambda}^{\mathrm{H}} \right) + \breve{\mathbf{N}}_{k} \notag \\
    & = \mathbf{\Theta}_{k}^{\mathrm{H}} \mathbf{H}_{\mathrm{RIS}}^{k} + \breve{\mathbf{N}}_{k},
    \label{eq19}
\end{align}
where \( \mathbf{H}_{\mathrm{RIS}}^{k} \triangleq \left( \mathbf{A}_{M,k}^\mathrm{*} \otimes \mathbf{A}_{M} \right) \left( \boldsymbol{\beta}_{k}^\mathrm{*} \otimes \mathbf{\Lambda}^{\mathrm{H}} \right) \). By exploiting the structure of \(\mathbf{H}_{\mathrm{RIS}}^{1}\) associated with user~1, we extract the \(r\)-th column of the equivalent measurement matrix \(\breve{\mathbf{Y}}_{1}\), denoted as \(\breve{\mathbf{y}}_{1,r}\), which is given by
\begin{align}
    \breve{\mathbf{y}}_{1,r} 
    = & \mathbf{\Theta}_{1}^{\mathrm{H}} \left( \mathbf{A}_{M,1}^\mathrm{*} \otimes \mathbf{A}_{M} \right)  \left[ \left( \boldsymbol{\beta}_{1}^\mathrm{*} \otimes \mathbf{\Lambda}^{\mathrm{H}} \right) \right]_{:,r} + \mathbf{\breve{n}}_{1,r} \notag \\ 
    = & \mathbf{\Theta}_{1}^{\mathrm{H}} \left( \mathbf{A}_{M,1}^\mathrm{*} \otimes \mathbf{a}_{M} \left( \omega_{r}, \mu_{r} \right) \right) \left( {\alpha}_{r}^\mathrm{*} \boldsymbol{\beta}_{1}^\mathrm{*} \right) + \mathbf{\breve{n}}_{1,r} ,
    \label{eq20}
\end{align}
where \(\mathbf{A}_{M,1}^\mathrm{*} \otimes \mathbf{a}_{M} \left( \omega_{r}, \mu_{r} \right) = \left[ \right. \mathbf{a}_{M}^\mathrm{*} \left( \varphi_{1,1},  \theta_{1,1} \right) \otimes \mathbf{a}_{M} \left( \omega_{r}, \mu_{r} \right)\) \(, \cdots, \mathbf{a}_{M}^\mathrm{*} \left( \varphi_{1,J_{1}}, \theta_{1,J_{1}} \right) \otimes \mathbf{a}_{M} \left( \omega_{r}, \mu_{r} \right) \left. \right] \in \mathbb{C}^{M^2\times J_{1}}\) and \(\mathbf{\breve{n}}_{1,r}\) denotes the \(r\)-th column of \(\mathbf{\breve{N}}_{1}\). This result is derived by observing that \(\left[ \left( \boldsymbol{\beta}_{1}^\mathrm{*} \otimes \mathbf{\Lambda}^{\mathrm{H}} \right) \right]_{:,r}\) contains non-zero elements at every \(r\)-th position within blocks of \(L\) rows, forming the sequence \(\{\alpha_r^* \beta_{1,j}^*\}_{j=1}^{J_1}\). This structure allows the extraction and concatenation of the \(r\)-th column from each block of \(L\) columns of \( \mathbf{A}_{M,1}^\mathrm{*} \otimes \mathbf{A}_{M}\), resulting in the term \(\mathbf{h}_{\mathrm{RIS},r}^{1} \triangleq \left( \mathbf{A}_{M,1}^\mathrm{*} \otimes \mathbf{a}_{M} \left( \omega_{r}, \mu_{r} \right) \right) \left( {\alpha}_{r}^\mathrm{*} \boldsymbol{\beta}_{1}^\mathrm{*} \right)\), which corresponds to the \(r\)-th column of \(\mathbf{H}_{\mathrm{RIS}}^{1}\).

To estimate \(\mathbf{h}_{\mathrm{RIS},r}^{1}\),  Eq.~\eqref{eq20} can be approximated by using virtual angular domain (VAD) representation as 
\begin{align}
    \breve{\mathbf{y}}_{1,r} 
    = \mathbf{\Theta}_{1}^{\mathrm{H}} \left( \tilde{\mathbf{A}}_{M,1}^{\mathrm{*}} \otimes \tilde{\mathbf{A}}_{M} \right) \mathbf{b}_{1,r} + \mathbf{\breve{n}}_{1,r} ,
    \label{eq21}
\end{align}
where \(\tilde{\mathbf{A}}_{M,1} \in \mathbb{C}^{M \times D_1}\) and \(\tilde{\mathbf{A}}_{M} \in \mathbb{C}^{M \times D_M}\) denote overcomplete dictionary matrices. The columns of \(\tilde{\mathbf{A}}_{M,1}^{\mathrm{*}} \otimes \tilde{\mathbf{A}}_{M}\) contain possible values of \(\{\mathbf{a}_{M}(\varphi_{1,j}, \theta_{1,j})^{\mathrm{*}} \otimes \mathbf{a}_{M}(\omega_{r}, \mu_{r})\}_{j=1}^{J_1}\), evaluated over the discretized angle domains. As an example, \(\tilde{\mathbf{A}}_{M}\) is constructed as
\begin{subequations}
    \begin{align}
    \tilde{\mathbf{A}}_{M} & = \tilde{\mathbf{A}}_{M_v} \otimes \tilde{\mathbf{A}}_{M_h} ,
    \label{eq22a} \\
    \left[ \tilde{\mathbf{A}}_{M_v} \right]_{:,g_v} & =  \mathbf{a}_{M_{v}} \left( \left(-1 + \frac{2}{D_v}g_v \right) \frac{d_{\mathrm{RIS}}}{\lambda_{c}}\right) ,
    \label{eq22b} \\
    \left[ \tilde{\mathbf{A}}_{M_h} \right]_{:,g_h} & =  \mathbf{a}_{M_{h}} \left( \left(-1 + \frac{2}{D_h}g_h \right) \frac{d_{\mathrm{RIS}}}{\lambda_{c}}\right) ,
    \label{eq22c}
    \end{align}   
\end{subequations}
where \(g_v = 0, 1, \cdots, D_v - 1\) (with \(D_v \ge M_v\)) and \(g_h = 0, 1, \cdots, D_h - 1\) (with \(D_h \ge M_h\)), and \(D = D_v \times D_h\). The atoms in \(\tilde{\mathbf{A}}_{M_v}\) and \(\tilde{\mathbf{A}}_{M_h}\) span the intervals \( \left[ -\frac{d_{\mathrm{RIS}}}{\lambda_c}, \left(1 - \frac{2}{D_v} \right)\frac{d_{\mathrm{RIS}}}{\lambda_c} \right] \) and \( \left[ -\frac{d_{\mathrm{RIS}}}{\lambda_c}, \left(1 - \frac{2}{D_h} \right) \frac{d_{\mathrm{RIS}}}{\lambda_c} \right] \), with resolutions \(2/D_v\) and \(2/D_h\), respectively. Furthermore, \(\mathbf{b}_{1,r} \in \mathbb{C}^{D_M D_1 \times 1}\) is a sparse vector with \(J_1\) non-zero entries corresponding to the cascaded channel path gains \(\{\alpha_r^* \beta_{1,j}^*\}_{j=1}^{J_1}\)\footnote{The number of propagation paths between user~1 and the RIS, denoted by \(J_1\), determines the sparsity level in the recovery problems associated with Eqs.~\eqref{eq21} and~\eqref{eq27}. In Stage~II of the proposed protocol, OMP is used as the recovery algorithm. In this case, the stopping criterion is based on the residual power. Specifically, the algorithm terminates when the residual energy falls below a predefined threshold. The number of iterations is therefore treated as an estimate of \(J_1\).}. To ensure optimal CS performance, the MC-aware RIS phase shift training matrix \(\mathbf{\Theta}_{1}\) must be designed such that the columns of the equivalent dictionary \(\mathbf{\Theta}_{1}^{\mathrm{H}} \left( \tilde{\mathbf{A}}_{M,1}^{*} \otimes \tilde{\mathbf{A}}_{M} \right)\) are orthogonal. Designing the RIS phase shift training matrix simultaneously addresses the MC effect. The detailed Riemannian manifold optimization framework is discussed in Section~\ref{RIS}.

Using CS, the estimated value \(\widehat{\mathbf{h}}_{\mathrm{RIS},r}^{1}\) is obtained according to Eq.~\eqref{eq20}. The other columns of \(\mathbf{H}_{\mathrm{RIS}}^{1}\), denoted as \(\{\mathbf{h}_{\mathrm{RIS},l}^{1}\}_{l \ne r}^{L}\), can be estimated by recovering the AoAs at the RIS of user~1, i.e., the matrix \(\mathbf{A}_{M,1}\), thereby reducing computational complexity. Specifically, the term \(\mathbf{a}_{M}^{*}(\varphi_{1,j}, \theta_{1,j}) \otimes \mathbf{a}_{M}(\omega_{r}, \mu_{r}) = \left( \mathbf{a}_{M_v}^{*}(\varphi_{1,j}) \otimes \mathbf{a}_{M_h}^{*}(\theta_{1,j}) \right) \otimes \left( \mathbf{a}_{M_v}(\omega_{r}) \otimes \mathbf{a}_{M_h}(\mu_{r}) \right)\) represents a four-dimensional (4-D) angular product. A significant reduction in dimensionality can be achieved by applying a similar procedure as described in Eq.~\eqref{eq14}, thereby enabling accurate angle estimation with reduced computational burden. The dimension-reduced received signal matrix \(\mathbf{Y}_{\mathrm{AoA}_h}^1 \in \mathbb{C}^{M_h \times M M_v}\) is obtained via element extraction, as defined by
\begin{align}
    \left[ \mathbf{Y}_{\mathrm{AoA}_h}^1 \right]_{m_h, (m_v-1)M+m} = \left[ \mathbf{h}_{\mathrm{RIS},r}^{1} \right]_{i_h(m_h, m_v, m)} , 
    \label{eq23}
\end{align}
where \(i_h(m_h, m_v, m) = (m_h - 1)M + (m_v - 1)M M_h + m\) is a indexing function, and the indices \(m_h\), \(m_v\), and \(m\) take values from the sets \(\{1,\cdots,M_h\}\), \(\{1,\cdots,M_v\}\), and \(\{1,\cdots,M\}\), respectively. By omitting the row index indicator \(m_h\), the element-wise concatenated result can be expressed as
\begin{align}
    & \left[ \mathbf{Y}_{\mathrm{AoA}_h}^1 \right]_{:, (m_v-1)M+m} = \notag \\ & \left[ \kappa_h  \left(1,m,m_v \right) \mathbf{a}_{M_h}^\mathrm{*}(\theta_{1,1}), \cdots, \kappa_h \left(J_1,m,m_v \right) \mathbf{a}_{M_h}^\mathrm{*}(\theta_{1,J_1}) \right] \left( \mathbf{\alpha}_{r}^\mathrm{*} \boldsymbol{\beta}_{1}^\mathrm{*} \right) \notag \\
    & = \mathbf{A}_{M_h,1}^\mathrm{*} \mathrm{Diag} \left(\kappa_h \left(1,m,m_v \right), \cdots, \kappa_h \left(J_1,m,m_v \right) \right) \left( \mathbf{\alpha}_{r}^\mathrm{*} \boldsymbol{\beta}_{1}^\mathrm{*} \right) ,
    \label{eq24} 
\end{align}
where \(\frac{ln\left( \kappa_h \left(j,m,m_v \right) \right)}{-i 2\pi} =  \left( m-\lfloor \frac{m-1}{M_h}\rfloor M_h-1\right) \mu_{l} + \lfloor \frac{m-1}{M_h}\rfloor \) \( \omega_{l} - (m_v-1)\varphi_{1,j}\) and \(\mathbf{A}_{M_h,1} \triangleq \left[ \mathbf{a}_{M_h}(\theta_{1,1}), \cdots, \mathbf{a}_{M_h}(\theta_{1,J_1}) \right]\) denotes the horizontal steering matrix for the \(y\)-axis dimension of the UPA corresponding to the AoA at the RIS of user~1. After element-wise concatenation, a formulation suitable for sparse signal recovery is obtained. Analogously, the dimension-reduced received signal in the vertical dimension can also be derived via element extraction as follows
\begin{align}
    \left[ \mathbf{Y}_{\mathrm{AoA}_v}^1 \right]_{m_v, m^*} = \left[ \mathbf{h}_{\mathrm{RIS},r}^{1} \right]_{i_v \left(m_v, m*\right)} ,
    \label{eq25} 
\end{align}
where \(i_v(m_v, m^*)= (m_v-1)MM_h+m^*\) denotes the indexing function, and \(m^*\) takes values from the set \(\{1, \cdots, MM_h\}\). Then, by omitting the row index indicator \(m_v\) , the element-wise concatenated result can be obtained as:
\begin{align}
    & \left[ \mathbf{Y}_{\mathrm{AoA}_v}^1 \right]_{:, m^*} = \notag \\
    & \left[ \kappa_v  \left(1,m^* \right) \mathbf{a}_{M_v}^\mathrm{*}(\varphi_{1,1}), \cdots, \kappa_v \left(J_1,m^* \right) \mathbf{a}_{M_v}^\mathrm{*}(\varphi_{1,J_1}) \right] \left( \mathbf{\alpha}_{r}^\mathrm{*} \boldsymbol{\beta}_{1}^\mathrm{*} \right) \notag \\
    & =\mathbf{A}_{M_v,1}^\mathrm{*} \mathrm{Diag} \left(\kappa_v \left(1,m^* \right), \cdots, \kappa_v \left(J_1,m^* \right) \right) \left( \mathbf{\alpha}_{r}^\mathrm{*} \boldsymbol{\beta}_{1}^\mathrm{*} \right) ,
    \label{eq26} 
\end{align}
where \(\frac{ln \left(\kappa_v \left(j,m^* \right) \right)}{-i 2\pi} = \left( m^*- \lfloor \frac{m^*-\lfloor \frac{m^*-1}{M} \rfloor M-1}{M_h}  \rfloor M_h-1\right) \mu_{l} + \lfloor \frac{m^*-\lfloor \frac{m^*-1}{M} \rfloor M-1}{M_h}  \rfloor \omega_{l} - \lfloor \frac{m^*-1}{M} \rfloor \theta_{1,j}\) and \(\mathbf{A}_{M_v,1} \triangleq \left[ \mathbf{a}_{M_v}(\varphi_{1,1}), \cdots, \mathbf{a}_{M_v}(\varphi_{1,J_1}) \right]\) denotes the corresponding vertical steering matrix. Remarkably, Eqs.~\eqref{eq24} and~\eqref{eq26} are compatible with both subspace-based and CS techniques. In this work, we adopt the CS method outlined in Algorithm~\ref{alg2}. After executing the respective angle estimation procedures, \(\mathbf{A}_{M,1}\) can be obtained as \(\widehat{\mathbf{A}}_{M,1} = \widehat{\mathbf{A}}_{M_v,1} \diamond \widehat{\mathbf{A}}_{M_h,1}\). The other columns of \(\mathbf{H}_{\mathrm{RIS}}^{1}\), i.e., \(\{\mathbf{h}_{\mathrm{RIS},l}^{1}\}_{l \ne r}^{L}\), can then be estimated via a \(J_1\)-sparse signal recovery problem, thereby reducing computational complexity:
\begin{align}
    \breve{\mathbf{y}}_{1,l} 
    = \mathbf{\Theta}_{1}^{\mathrm{H}} \left( \widehat{\mathbf{A}}_{M,1}^{\mathrm{*}} \otimes \tilde{\mathbf{A}}_{M} \right) \mathbf{b}_{1,l} + \breve{\mathbf{n}}_{1,l} ,
    \label{eq27}
\end{align}
where \(\tilde{\mathbf{A}}_{M}\) is defined in Eq.~\eqref{eq22a}-\eqref{eq22c}. Therefore, the other columns \(\widehat{\mathbf{h}}_{\mathrm{RIS},l}^{1}\), for \(1 \le l \le \widehat{L}, l \ne r\), can be obtained. Finally, the estimated cascaded channel for user~1 is expressed as
\begin{align}
    \widehat{\mathbf{G}}_{1} = \widehat{\mathbf{A}}_{N} \left[ \widehat{\mathbf{h}}_{\mathrm{RIS},1}^{1}, \cdots, \widehat{\mathbf{h}}_{\mathrm{RIS},\widehat{L}}^{1} \right]^\mathrm{H}.
    \label{eq28}
\end{align}
The overall estimation of \(\mathbf{G}_{1}\) is summarized in Algorithm~\ref{alg2}.
\begin{algorithm}
\caption{Estimation of MC-aware cascaded channel for Typical User}
\label{alg2}
\begin{algorithmic}[1]  
\REQUIRE \(\mathbf{Y}_{1}\)
\STATE Obtain estimated number of paths in RIS-BS channel \(\widehat{L}\) and common AoA steering matrix \(\widehat{\mathbf{A}}_{N}\) from Algorithm~\ref{alg1}.
\STATE Compute the equivalent measurement matrix \(\left[ \breve{\mathbf{y}}_{1,1}, \cdots, \breve{\mathbf{y}}_{1,\widehat{L}} \right] = \left( \frac{1}{N \sqrt{p}} \widehat{\mathbf{A}}_{N}^{\mathrm{H}} \mathbf{Y}_{1} \right)^{\mathrm{H}}\).
\STATE Determine the reference column index \(r\) as:
\begin{align}
    r = \arg\max_{1 \le i \le \widehat{L}} \left\| \breve{\mathbf{y}}_{1,i} \right\|_2^2.
    \label{eq29}
\end{align}
\STATE Construct the dictionary \(\mathbf{D}_{1} = \mathbf{\Theta}_{1}^{\mathrm{H}} \left(\tilde{\mathbf{A}}_{M,1}^{\mathrm{*}} \otimes \tilde{\mathbf{A}}_{M} \right)\).
\STATE Estimate \(\widehat{\mathbf{h}}_{\mathrm{RIS},r}^{1}\) from \(\breve{\mathbf{y}}_{1,r}\) using \(\mathbf{D}_{1}\) and the OMP algorithm with Eq.~\eqref{eq21}.
\STATE Estimate \(\widehat{\mathbf{A}}_{M_h,1}\) from \(\widehat{\mathbf{h}}_{\mathrm{RIS},r}^{1}\) using Eq.~\eqref{eq23} via element extraction and the OMP algorithm with dictionary \(\mathbf{\Xi}_{1,h} = \tilde{\mathbf{A}}_{M_h,1}^{\mathrm{*}}\), constructed similarly to Eq.~\eqref{eq22c}.
\STATE Estimate \(\widehat{\mathbf{A}}_{M_v,1}\) from \(\widehat{\mathbf{h}}_{\mathrm{RIS},r}^{1}\) using Eq.~\eqref{eq25} via element extraction and the OMP algorithm with dictionary \(\mathbf{\Xi}_{1,v} = \tilde{\mathbf{A}}_{M_v,1}^{\mathrm{*}}\), constructed similarly to Eq.~\eqref{eq22b}.
\STATE Recover the AoA steering matrix at the RIS of user~1 as \(\widehat{\mathbf{A}}_{M,1} = \widehat{\mathbf{A}}_{M_v,1} \diamond \widehat{\mathbf{A}}_{M_h,1}\).
\STATE Construct the dictionary \(\mathbf{\Xi}_{1} = \mathbf{\Theta}_{1}^{\mathrm{H}} \left(\widehat{\mathbf{A}}_{M,1}^{\mathrm{*}} \otimes \tilde{\mathbf{A}}_{M} \right)\).
\FOR{\(1 \leq l \leq \widehat{L},\ l \ne r\)}
\STATE Estimate \(\widehat{\mathbf{h}}_{\mathrm{RIS},l}^{1}\) from \(\breve{\mathbf{y}}_{1,l}\) using \(\mathbf{\Xi}_{1}\) and the OMP algorithm with Eq.~\eqref{eq27}.
\ENDFOR
\ENSURE \(\widehat{\mathbf{G}}_{1} = \widehat{\mathbf{A}}_{N} \left[ \widehat{\mathbf{h}}_{\mathrm{RIS},1}^{1}, \cdots, \widehat{\mathbf{h}}_{\mathrm{RIS},\widehat{L}}^{1} \right]^\mathrm{H}\).
\end{algorithmic}
\end{algorithm}

Notably, the OMP algorithm facilitates the extraction of the set \(\{ \widehat{\mathbf{A}}_{M,1}^\mathrm{*} \otimes \widehat{\mathbf{a}}_{M}(\omega_{l}, \mu_{l}) \}_{l=1}^{\widehat{L}}\) from the equivalent dictionaries \(\mathbf{D}_{1}\) and \(\mathbf{\Xi}_{1}\). By leveraging the properties of the Kronecker product and the structure in Eq.~\eqref{eq20}, these components are concatenated to construct the matrix \( \widehat{\mathbf{A}}_{M,1}^\mathrm{*} \otimes \widehat{\mathbf{A}}_{M} \). Considering that \( \left[ \mathbf{A}_{M,1}^\mathrm{*} \right]_{1,1} = 1 \), the common AoD steering matrix at the RIS can be directly obtained as \( \widehat{\mathbf{A}}_{M} = \left[ \widehat{\mathbf{A}}_{M,1}^\mathrm{*} \otimes \widehat{\mathbf{A}}_{M} \right]_{1:M,1:\widehat{L}} \).

\subsection{Stage~III: Channel Estimation for Other Users}

In this subsection, we exploit the fact that all users share the common RIS-BS channel to reduce the pilot overhead in channel estimation. Building upon the common AoA steering matrix \(\widehat{\mathbf{A}}_N\) at the BS, as derived in Section~\ref{AoABS}, the equivalent measurement matrix for the \(k\)-th user can be constructed according to Eq.~\eqref{eq19}. This results in a measurement model analogous to that in Eq.~\eqref{eq20}, expressed as \(\breve{\mathbf{y}}_{k,r} = \mathbf{\Theta}_{k}^{\mathrm{H}} \left( \mathbf{A}_{M,k}^\mathrm{*} \otimes \mathbf{a}_{M}(\omega_{r}, \mu_{r}) \right) \left( \mathbf{\alpha}_{r}^\mathrm{*} \boldsymbol{\beta}_{k}^\mathrm{*} \right) + \mathbf{\breve{n}}_{k,r}\). Leveraging the common AoD steering matrix \(\widehat{\mathbf{A}}_M\) obtained in Section~\ref{AoDRIS}, the \(J_k\)-sparse signal recovery problem for other users can be reformulated to reduce computational complexity as follows
\begin{align}
    \breve{\mathbf{y}}_{k,r} 
    = \mathbf{\Theta}_{k}^{\mathrm{H}} \left( \tilde{\mathbf{A}}_{M,k}^{\mathrm{*}} \otimes \widehat{\mathbf{A}}_{M} \right) \mathbf{b}_{k,r} + \mathbf{\breve{n}}_{k,r} ,
    \label{eq30}
\end{align}
where the construction of \(\tilde{\mathbf{A}}_{M,k}\) follows the same procedure as in Eq.~\eqref{eq22a}-\eqref{eq22c}. Similarly, after obtaining the estimate \(\widehat{\mathbf{h}}_{\mathrm{RIS},r}^{k}\) via Eq.~\eqref{eq30}, we exploit its 4-D angular product structure, \(\mathbf{a}_{M}^{*}(\varphi_{k,j}, \theta_{k,j}) \otimes \mathbf{a}_{M}(\omega_{r}, \mu_{r}) = \left( \mathbf{a}_{M_v}^{*}(\varphi_{k,j}) \otimes \mathbf{a}_{M_h}^{*}(\theta_{k,j}) \right) \otimes \left( \mathbf{a}_{M_v}(\omega_{r}) \otimes \mathbf{a}_{M_h}(\mu_{r}) \right)\) to recover the horizontal and vertical AoA steering matrices \(\mathbf{A}_{M_h,k}\) and \(\mathbf{A}_{M_v,k}\) via Eqs.~\eqref{eq23} and~\eqref{eq25}, respectively. We then derive the full AoA steering matrix at the RIS of user~\(k\) as \(\widehat{\mathbf{A}}_{M,k} = \widehat{\mathbf{A}}_{M_v,k} \diamond \widehat{\mathbf{A}}_{M_h,k}\), which reduces the computational complexity associated with estimating the remaining columns of \(\mathbf{H}_{\mathrm{RIS}}^{k}\), i.e., \(\{\widehat{\mathbf{h}}_{\mathrm{RIS},l}^{k}\}_{l \ne r}^{L}\). The specific pilot overhead and computational complexity reduction achieved through the above operations will be discussed in Section~\ref{analysis}.

\subsection{MC-unaware Channel Estimation}\label{unaware}

This subsection presents a channel estimation scheme tailored to the conventional RIS-aided cascaded channel model as described in Eq.~\eqref{eq8}. In Section~\ref{simulation}, the scheme is applied to received signals under the MC effect and compared with the proposed and existing MC-aware algorithms. This comparative analysis rigorously characterizes the impact of MC on estimation performance while highlighting the advantages of the proposed MC-aware estimation protocol.

Similar to Eq.~\eqref{eq10}, the received signal that ignores the MC effect can be expressed as
\begin{align}
    \mathbf{y}_{\mathrm{cv},k}(t) 
    & = \sqrt{p} \mathbf{h}_{\mathrm{cv},k} + \mathbf{n}_{\mathrm{cv},k}(t) \notag \\
    & = \sqrt{p} \mathbf{A}_{N} \mathbf{\Lambda} \mathbf{A}_{M}^{\mathrm{H}} \mathbf{\Gamma} \mathbf{A}_{M,k} \boldsymbol{\beta}_{k} + \mathbf{n}_{\mathrm{cv},k}(t) \notag \\
    & = \sqrt{p} \mathbf{A}_{N} \mathbf{\Lambda} \mathbf{A}_{M}^{\mathrm{H}} \mathrm{Diag} \left( \mathbf{A}_{M,k} \boldsymbol{\beta}_{k}\right) \boldsymbol{\gamma} + \mathbf{n}_{\mathrm{cv},k}(t) ,
    \label{eq31}
\end{align}
where the diagonal structure of the RIS response enables \( \mathbf{\Gamma} = \mathrm{Diag}(\boldsymbol{\gamma})\) the above reformulation. In contrast, when the MC effect is incorporated, the RIS response is no longer diagonal and be modeled by \(\left(\mathbf{\Gamma}^{-1}-\mathbf{S}\right)^{-1}\), which fundamentally alters the cascaded channel structure and requires additional signal processing. Similar to Eq.~\eqref{eq12}, stacking the \(\tau_k\) received signal vectors across time slots yields \(\mathbf{Y}_{\mathrm{cv},k} \in \mathbb{C}^{N \times \tau_k}\) as follows
\begin{align}
    \mathbf{Y}_{\mathrm{cv},k} 
    & = \left[ \mathbf{y}_{\mathrm{cv},k}(1), \cdots, \mathbf{y}_{\mathrm{cv},k}(\tau_k) \right] \notag \\ 
    & \triangleq \sqrt{p} \mathbf{G}_{\mathrm{cv},k} \mathbf{\Theta}_{\mathrm{cv},k} + {\mathbf{N}}_{\mathrm{cv},k},
    \label{eq32}
\end{align}
where \(\mathbf{G}_{\mathrm{cv},k} = \mathbf{A}_{N} \mathbf{\Lambda} \mathbf{A}_{M}^{\mathrm{H}} \mathrm{Diag} \left( \mathbf{A}_{M,k} \boldsymbol{\beta}_{k} \right)\) defines the MC-unaware cascaded user–RIS–BS channel for user \(k\). In addition, \(\mathbf{\Theta}_{\mathrm{cv},k} = \left[ \boldsymbol{\gamma}_{1}, \cdots, \boldsymbol{\gamma}_{\tau_k} \right] \in \mathbb{C}^{M \times \tau_k}\) denotes the RIS phase shift training matrix for user \(k\), and \(\mathbf{N}_{\mathrm{cv},k} = \left[ \mathbf{n}_{\mathrm{cv},k}(1), \cdots, \mathbf{n}_{\mathrm{cv},k} \left( \tau_{k} \right) \right] \in \mathbb{C}^{N \times \tau_{k}}\) represents the corresponding noise matrix.

By examining the structure of \(\mathbf{G}_{\mathrm{cv},k}\), it can be observed that the estimation procedure for the common AoA at the BS is identical to that described in Section~\ref{AoABS}. After eliminating the common AoA influence, the equivalent MC-unaware measurement signal can be obtained as
\begin{align}
    \breve{\mathbf{Y}}_{\mathrm{cv},k}  
    & \triangleq \left( \frac{1}{N \sqrt{p}} \widehat{\mathbf{A}}_{N}^{\mathrm{H}} \mathbf{Y}_{\mathrm{cv},k} \right)^{\mathrm{H}} \notag \\
    & = \mathbf{\Theta}_{\mathrm{cv},k}^{\mathrm{H}} \mathrm{Diag} \left( \mathbf{h}_{k}^{*} \right) \mathbf{A}_{M} \mathbf{\Lambda}^{*} + \mathbf{\breve{N}}_{\mathrm{cv},k} \notag \\
    & = \mathbf{\Theta}_{\mathrm{cv},k}^{\mathrm{H}} \mathbf{H}_{\mathrm{RIS}}^{\mathrm{cv},k} + \mathbf{\breve{N}}_{\mathrm{cv},k},
    \label{eq33}
\end{align}
where \( \mathbf{H}_{\mathrm{RIS}}^{\mathrm{cv},k} \triangleq \mathrm{Diag} \left( \mathbf{h}_{k}^{*} \right) \mathbf{A}_{M} \mathbf{\Lambda}^{*} \) and \(\mathbf{\breve{N}}_{\mathrm{cv},k}\) is the corresponding noise. By exploiting the structure of \(\mathbf{H}_{\mathrm{RIS}}^{\mathrm{cv},k}\), we obtain
\begin{align}
    \mathbf{H}_{\mathrm{RIS}}^{\mathrm{cv},k} 
    & = \mathbf{h}_{k}^{*} \bullet \left( \mathbf{A}_{M} \mathbf{\Lambda}^{*} \right) = \left( \mathbf{A}_{M,k} \boldsymbol{\beta}_k \right)^{*} \bullet \left( \mathbf{A}_{M} \mathbf{\Lambda}^{*} \right) \notag \\
    & = \left( \mathbf{A}_{M,k}^{*} \bullet \mathbf{A}_{M} \right) \left( \boldsymbol{\beta}_{k}^{*} \otimes \mathbf{\Lambda}^{*} \right),
    \label{eq34}
\end{align}
where \(\mathbf{A}_{M,k}^{*} \bullet \mathbf{A}_{M} = \left[\right. \mathbf{a}_M(\omega_1 - \varphi_{k,1}, \mu_1 - \theta_{k,1}), \cdots, \mathbf{a}_M(\omega_{L}\) \( - \varphi_{k,J_k}, \mu_{L} - \theta_{k,J_k}) \left.\right] \in \mathbb{C}^{M \times LJ_k}\). The final equality follows from the identity \( \left( \mathbf{A} \bullet \mathbf{B} \right) \left( \mathbf{C} \otimes \mathbf{D} \right) = \left( \mathbf{A} \mathbf{C}) \bullet \left( \mathbf{B} \mathbf{D} \right) \right) \)~\cite{ref24}. Similar to Eq.~\eqref{eq20}, channel estimation can be performed via column extraction as follows
\begin{align}
    \mathring{\mathbf{y}}_{k,l} 
    = & \mathbf{\Theta}_{\mathrm{cv},k}^{\mathrm{H}} \mathrm{Diag} \left( \mathbf{h}_{k}^{*} \right) \left[ \mathbf{A}_{M} \mathbf{\Lambda}^{*} \right]_{:,l} + \mathring{\mathbf{n}}_{k,l} \notag \\ 
    = & \mathbf{\Theta}_{\mathrm{cv},k}^{\mathrm{H}} \mathbf{h}_{k}^{*} \bullet \left( \alpha_{l}^{*} \mathbf{a}_{M} \left( \omega_{l}, \mu_{l} \right) \right) + \mathring{\mathbf{n}}_{k,l} \notag \\
    = & \mathbf{\Theta}_{\mathrm{cv},k}^{\mathrm{H}} \left( \mathbf{A}_{M,k}^{*} \bullet \mathbf{a}_{M} \left( \omega_{l}, \mu_{l} \right) \right)\alpha_{l}^{*} \boldsymbol{\beta}_k + \mathring{\mathbf{n}}_{k,l} ,
    \label{eq35}
\end{align}
where \( \mathbf{A}_{M,k}^{*} \bullet \mathbf{a}_M(\omega_l, \mu_l) = 
\left[\right. \mathbf{a}_M(\omega_l - \varphi_{k,1}, \mu_l - \theta_{k,1}), \cdots, \)  \(\mathbf{a}_M(\omega_l - \varphi_{k,J_k}, \mu_l - \theta_{k,J_k})\left.\right] 
\in \mathbb{C}^{M \times J_k} \), and \( \mathring{\mathbf{n}}_{k,l} \) denotes the \( l \)-th column of \( \mathring{\mathbf{N}}_{\mathrm{cv},k} \). Similar to the discussion following Eq.~\eqref{eq20}, Eq.~\eqref{eq35} can also be directly derived from Eq.~\eqref{eq34} by exploiting the block structure induced by the Kronecker and transposed Khatri-Rao products. Note that \(\mathrm{Diag} \left( \mathbf{h}_{k}^{*} \right) \left[ \mathbf{A}_M \mathbf{\Lambda}^{*} \right]_{:,l}\) corresponds to the \(l\)-th column of \(\mathbf{H}_{\mathrm{RIS}}^{\mathrm{cv},k}\), denoted as \(\mathbf{h}_{\mathrm{RIS},l}^{\mathrm{cv},k}\). Given that all angles lie within the interval \(\left[-\frac{d_{\mathrm{RIS}}}{\lambda_c}, \frac{d_{\mathrm{RIS}}}{\lambda_c}\right]\), Eq.~\eqref{eq35} can be reformulated as a \(J_k\)-sparse signal recovery problem, which can be efficiently solved using CS techniques. By jointly processing all \(\widehat{L}\) columns for each of the \(K\) users, the complete set of \(\{ \mathbf{H}_{\mathrm{RIS}}^{\mathrm{cv},k} \}_{k=1}^{K}\) can be achieved. Consequently, the MC-unaware equivalent cascaded channel estimates for all users are obtained as \(\widehat{\mathbf{G}}_{\mathrm{cv},k} = \widehat{\mathbf{A}}_{N} \left[ \widehat{\mathbf{h}}_{\mathrm{RIS},1}^{\mathrm{cv},k}, \cdots, \widehat{\mathbf{h}}_{\mathrm{RIS},\widehat{L}}^{\mathrm{cv},k} \right]^\mathrm{H}\).

\subsection{Pilot Overhead and Computational Complexity Analysis}\label{analysis}

In this subsection, we analyze the pilot overhead and computational complexity of the proposed MC-aware three-stage channel estimation protocol. For simplicity, we assume that \( J_1 = \cdots = J_K = J \).

\begin{table*}[t]
\caption{Pilot Overhead and Computational Complexity Comparison\label{table1}}
\centering
\renewcommand{\arraystretch}{1.15}
\resizebox{\linewidth}{!}{%
\begin{tabular}{|c|c|c|}
        \hline
        \textbf{Algorithm} & \textbf{Proposed~1 / Proposed~2} & \textbf{Direct-OMP} \\
        \hline
        \multicolumn{3}{|c|}{\textbf{Pilot Overhead}} \\
        \hline
        User~1
        & $\tau_1 \geq \mathcal{O}(J_1 \log(D_MD_1)) \geq \mathcal{O}(J \log(M^2))$
        & \multirow{2}{*}{\makecell[c]{$\tilde{\tau}_k \geq \mathcal{O}(L^2J_k \log(D_ND_MD_k)/N)$ \\ $\geq \mathcal{O}(L^2J \log(M^2N)/N)$}} \\
        \cline{1-2}
        User~\(k\) (\(2 \leq k \leq K\))
        & $\tau_k \geq \mathcal{O}(J_k \log(L D_k)) \geq \mathcal{O}(J \log(ML))$
        & \\
        \hline
        \multicolumn{3}{|c|}{\textbf{Computational Complexity}} \\
        \hline
        Stage~I
        & $\mathcal{O}\!\left(N\!\left(N_h+N_v\right)\!\sum_{k=1}^{K}\tau_k + N_h^{3}+N_v^{3}\right)$
        & \multirow{3}{*}{$\mathcal{O}\!\left(L^{2}J ND_ND_M\sum_{k=1}^{K}\left(\tilde{\tau}_k D_k\right)\right)$} \\
        \cline{1-2}
        Stage~II
        & $\mathcal{O}\!\left(\tau_1D_MJ\!\left(D_1+(L-1)J\right) + J\!\left(M_hD_{1,h}+M_vD_{1,v}\right)\right)$
        & \\
        \cline{1-2}
        Stage~III
        & $\mathcal{O}\!\left(\sum_{k=2}^{K}\Big(\tau_kLJ\!\left(D_k+(L-1)J\right) + J\!\left(M_hD_{k,h}+M_vD_{k,v}\right)\Big)\right)$
        & \\
        \hline
    \end{tabular}
}
\end{table*}

\subsubsection{Pilot Overhead Analysis}\label{pilot}

The pilot overhead \( \tau_1 \) for the typical user and \( \tau_k \) for user~\(k\) (\(2 \leq k \leq K\)) are associated with different stages of the proposed protocol. Specifically, \( \tau_1 \) governs the measurement dimension for the \(J_1\)-sparse signal recovery problem in Eq.~\eqref{eq21} in Stage~II, while \( \tau_k \) determines the required pilot overhead for the \(J_k\)-sparse signal recovery problem in Eq.~\eqref{eq30} in Stage~III. Both \( \tau_1 \) and \( \tau_k \) jointly influence the common AoA estimation in Stage~I.

According to compressive sensing theory~\cite{ref23}, recovering a \( l \)-sparse complex signal of dimension \( n \) requires the number of measurements \(m\) on the order of \(\mathcal{O}(l \log n)\). For the typical user, the sensing matrix in Eq.~\eqref{eq21} is given by \( \mathbf{\Theta}_1^{\mathrm{H}} \left( \tilde{\mathbf{A}}_{M,1}^{*} \otimes \tilde{\mathbf{A}}_M \right) \in \mathbb{C}^{\tau_1 \times D_M D_1} \), indicating that the required pilot overhead satisfies \( \tau_1 \geq \mathcal{O}(J_1 \log(D_MD_1)) \geq \mathcal{O}(J \log(M^2)) \). For other users, the sensing matrix in Eq.~\eqref{eq30} is \( \mathbf{\Theta}_k^{\mathrm{H}} \left( \tilde{\mathbf{A}}_{M,k}^{*} \otimes \widehat{\mathbf{A}}_M \right) \in \mathbb{C}^{\tau_k \times L D_k} \), and the required pilot overhead satisfies \( \tau_k \geq \mathcal{O}(J_k \log(L D_k)) \geq \mathcal{O}(J \log(ML))  \). 

\subsubsection{Computational Complexity Analysis}

For estimating the common AoA at the BS in Stage~I, as described in Algorithm~\ref{alg1}, the computational complexity primarily arises from the MLE of the covariance matrix and the eigenvalue decomposition, with respective complexities of \(\mathcal{O}\left(N_v N_h^2 \sum_{k=1}^{K} \tau_{k}\right) = \mathcal{O}\left(N N_h \sum_{k=1}^{K} \tau_{k}\right)\) and \(\mathcal{O}\left(N_h^3\right)\) in the horizontal dimension. The vertical dimension has an identical form. For estimating the remaining parameters of the typical user in Stage~II, as outlined in Algorithm~\ref{alg2}, the computational complexity primarily arises from the OMP-based method used for estimating \(\mathbf{h}_{\mathrm{RIS},r}^{1}\) in Step~5, \(\mathbf{A}_{M,1}\) in Steps~6 and~7, and \(\mathbf{h}_{\mathrm{RIS},l}^{1}\) (\(1 \leq l \leq \widehat{L},\ l \ne r\)) in Step~11. Since the dominant computational complexity of OMP is \(\mathcal{O}(mnl)\)~\cite{ref17}, where \(m\) denotes the number of measurements, \(n\) the length of the sparse signal with sparsity level \(l\), the computational complexities of these steps are \(\mathcal{O}(\tau_1 D_M D_1 J)\), \(\mathcal{O}(M_h D_{1,h} J)\), \(\mathcal{O}(M_v D_{1,v} J)\), and \(\mathcal{O}((L-1)\tau_1 D J^2)\), respectively. From the above analysis, it can be seen that estimating the vertical and horizontal directions separately transforms the complexity from a multiplicative form to an additive form.

Then, for estimating the remaining parameters of the other users in Stage~III, the computational complexity also arises from the OMP-based method used for estimating \(\mathbf{h}_{\mathrm{RIS},r}^{k}\), \(\mathbf{A}_{M,k}\), and \(\mathbf{h}_{\mathrm{RIS},l}^{k}\) (\(1 \leq l \leq \widehat{L},\ l \ne r\)). Similarly, for user~\(k\) (\(2 \leq k \leq K\)), the corresponding computational complexities of these steps are \(\mathcal{O}(\tau_k D_k L J)\), \(\mathcal{O}(M_h D_{k,h} J)\), \(\mathcal{O}(M_v D_{k,v} J)\), and \(\mathcal{O}((L-1)\tau_k L J^2)\), respectively. As observed above, estimating \(\widehat{\mathbf{A}}_{M}\) significantly reduces the computational complexity for other users, with the reduction specifically reflected in replacing \(D_M\) with \(L\). We summarize the pilot overhead and computational complexity in Table~\ref{table1} and compare them with those of the Direct-OMP algorithm, which is introduced in Section~\ref{simulation}. Here, \(D_N\) denotes the dimension of the overcomplete dictionary matrix associated with \(\mathbf{A}_{N}\).

\section{Phase Shift Matrix Design}\label{RIS}

The performance of the OMP-based channel estimation is strongly influenced by the orthogonality of the equivalent dictionary. To this end, we optimize the RIS phase shift training matrices to generate approximately orthogonal dictionaries. Specifically, \( \mathbf{\Theta}_{\mathrm{cv},k} \in \mathbb{C}^{M \times \tau_{k}}\) (\( \forall k \in \mathcal{K} \)) are designed to enhance the ability of OMP and mitigate the influence of MC effect to recover the sparse vectors \( \mathbf{b}_{k,l} \) in problems defined by Eqs.~\eqref{eq21} and~\eqref{eq30}. 

Our approach is inspired by the theoretical result in~\cite{ref25}, which states that successful recovery of the sparse signal \( \mathbf{b}_{k,l} \) via OMP is guaranteed when the following condition holds:
\begin{align}
    \| \mathbf{b}_{k,l} \|_0 \leq \frac{1}{2} \left( 1 + \frac{1}{\upsilon} \right),
    \label{eq36}
\end{align}
where \( \upsilon \) denotes the mutual coherence of equivalent dictionary \( \mathbf{D}_{k} = \mathbf{\Theta}_{\mathrm{mc},k}^{\mathrm{H}} \left( \tilde{\mathbf{A}}_{M,k}^{\mathrm{*}} \otimes \tilde{\mathbf{A}}_{M} \right) \in \mathbb{C}^{\tau_{k} \times DD_k}\), given by
\begin{align}
    \upsilon = \max_{i \ne j} \frac{|\mathbf{D}^{\mathrm{H}}_{k(:,i)} \mathbf{D}_{k(:,j)}|}{\| \mathbf{D}_{k(:,i)} \|_2 \| \mathbf{D}_{k(:,j)} \|_2}.
    \label{eq37}
\end{align}
The condition in Eq.~\eqref{eq37} indicates that the dictionary \( \mathbf{D}_{k} \) should exhibit low mutual coherence, i.e., its columns should be approximately orthogonal. This leads to the following optimization problem:
\begin{align}
    \min_{\mathbf{\Theta}_{\mathrm{cv},k}} & \left\|  \frac{1}{\tau_k} \mathbf{D}_{k}^{\mathrm{H}} \mathbf{D}_{k} - \mathbf{I}_{DD_k} \right\|_F^2 \notag \\
    \text{s.t.} \, & |\left[ \mathbf{\Theta}_{\mathrm{cv},k} \right]_{m,t}| = 1, 1 \leq m \leq M, 1 \leq t \leq \tau_k.
    \label{eq38}
\end{align}
The unconstrained version of Problem~\eqref{eq38} was previously studied in~\cite{ref26}, and the approach therein was extended to handle the unit-modulus constraint in~\cite{ref15}. Building on these works, a more concise formulation for the objective is adopted. Observe that
\begin{align}
    & \left\| \frac{1}{\tau_k} \mathbf{D}_{k}^{\mathrm{H}} \mathbf{D}_{k} - \mathbf{I}_{DD_k} \right\|_F^2 \notag \\
    = & \mathrm{tr} \left( \frac{1}{\tau_k^2} \mathbf{D}_{k}^{\mathrm{H}} \mathbf{D}_{k} \mathbf{D}_{k}^{\mathrm{H}} \mathbf{D}_{k} - \frac{2}{\tau_k} \mathbf{D}_{k}^{\mathrm{H}} \mathbf{D}_{k} + \mathbf{I}_{DD_k} \right) \notag \\
    = & \mathrm{tr} \left( \frac{1}{\tau_k^2} \mathbf{D}_{k} \mathbf{D}_{k}^{\mathrm{H}} \mathbf{D}_{k} \mathbf{D}_{k}^{\mathrm{H}} - \frac{2}{\tau_k} \mathbf{D}_{k} \mathbf{D}_{k}^{\mathrm{H}} + \mathbf{I}_{\tau_k} \right) + (DD_k - \tau_k) \notag \\
    = & \left\| \frac{1}{\tau_k} \mathbf{D}_{k} \mathbf{D}_{k}^{\mathrm{H}} - \mathbf{I}_{\tau_k} \right\|_F^2 + (DD_k - \tau_k).
    \label{eq39}
\end{align}

Considering the structure of \(\mathbf{D}_{k}\) and properties of the Kronecker product, Eq.~\eqref{eq39} can be simplified as:
\begin{align}
    \mathbf{D}_{k} \mathbf{D}_{k}^{\mathrm{H}} 
    & = \mathbf{\Theta}_{\mathrm{mc},k}^{\mathrm{H}} \left( \tilde{\mathbf{A}}_{M,k}^{\mathrm{*}} \otimes \tilde{\mathbf{A}}_{M} \right) \left( \tilde{\mathbf{A}}_{M,k}^{\mathrm{*}} \otimes \tilde{\mathbf{A}}_{M} \right)^{\mathrm{H}} \mathbf{\Theta}_{\mathrm{mc},k} \notag \\
    & = \mathbf{\Theta}_{\mathrm{mc},k}^{\mathrm{H}} \left( \tilde{\mathbf{A}}_{M,k}^{\mathrm{*}}\tilde{\mathbf{A}}_{M,k}^{\mathrm{T}} \otimes \tilde{\mathbf{A}}_{M} \tilde{\mathbf{A}}_{M}^{\mathrm{H}} \right) \mathbf{\Theta}_{\mathrm{mc},k}.
    \label{eq40}
\end{align}
Based on the over-complete dictionary construction in Eqs.~\eqref{eq22a}–\eqref{eq22c} and the asymptotic orthogonality of steering vector matrices established in~\cite{ref17}, we have:
\begin{align}
    \tilde{\mathbf{A}}_{M} \tilde{\mathbf{A}}_{M}^{\mathrm{H}} 
    & = \left( \tilde{\mathbf{A}}_{M_v} \otimes \tilde{\mathbf{A}}_{M_h} \right) \left( \tilde{\mathbf{A}}_{M_v} \otimes \tilde{\mathbf{A}}_{M_h} \right)^{\mathrm{H}} \notag \\
    & = \tilde{\mathbf{A}}_{M_v} \tilde{\mathbf{A}}_{M_v}^{\mathrm{H}} \otimes \tilde{\mathbf{A}}_{M_h} \tilde{\mathbf{A}}_{M_h}^{\mathrm{H}} \notag \\
    & \approx D_v \mathbf{I}_{M_v} \otimes D_h \mathbf{I}_{M_h} = D \mathbf{I}_{M}.
    \label{eq41}
\end{align}
Following the reasoning in Eq.~\eqref{eq41}, we also have \( \tilde{\mathbf{A}}_{M,k}^{*} \tilde{\mathbf{A}}_{M,k}^{\mathrm{T}} \approx D_k \mathbf{I}_{M} \). Substituting this and Eq.~\eqref{eq41} into Eq.~\eqref{eq40}, we obtain  \(\mathbf{D}_{k} \mathbf{D}_{k}^{\mathrm{H}} \approx \mathbf{\Theta}_{\mathrm{mc},k}^{\mathrm{H}} \left( D_k \mathbf{I}_{M} \otimes D \mathbf{I}_{M} \right) \mathbf{\Theta}_{\mathrm{mc},k} = D D_k \mathbf{\Theta}_{\mathrm{mc},k}^{\mathrm{H}} \mathbf{\Theta}_{\mathrm{mc},k}\). Hence, by applying the same derivation steps as in Eq.~\eqref{eq39}, the original optimization problem in Eq.~\eqref{eq38} simplifies to

\begin{align}
    \min_{\mathbf{\Theta}_{\mathrm{cv},k}} \, & \left\|  \frac{1}{\tau_k}\mathbf{\Theta}_{\mathrm{mc},k} \mathbf{\Theta}_{\mathrm{mc},k}^{\mathrm{H}} - \mathbf{I}_{M^2} \right\|_F^2 \notag \\
    \text{s.t.} \, & |\left[ \mathbf{\Theta}_{\mathrm{cv},k} \right]_{m,t}| = 1, 1 \leq m \leq M, \; 1 \leq t \leq \tau_k.
    \label{eq42}
\end{align}
This reformulation not only ensures the approximate orthogonality of the equivalent dictionary, but also facilitates the application of Eq.~\eqref{eq19} within subspace-based algorithms under reduced pilot overhead\footnote{Specifically, directly applying subspace-based algorithms would require left-multiplying Eq.~\eqref{eq19} by the pseudo-inverse of \(\mathbf{\Theta}_{\mathrm{mc},k}^{\mathrm{H}} \in \mathbb{C}^{\tau_k \times M^2}\), which necessitates \(\tau_k \ge M^2\) and leads to excessive pilot overhead. In contrast, the proposed design in Problem~\eqref{eq42} enables a standard signal model by left-multiplying Eq.~\eqref{eq19} with \(\mathbf{\Theta}_{\mathrm{mc},k}\) itself, while maintaining low pilot overhead.}.

To tackle the non-convex optimization problem in Eq.~\eqref{eq42} subject to unit-modulus constraints on complex variables, an efficient Riemannian manifold optimization framework is developed. Let \( f = \left\| \frac{1}{\tau_k}\mathbf{\Theta}_{\mathrm{mc},k} \mathbf{\Theta}_{\mathrm{mc},k}^{\mathrm{H}} - \mathbf{I}_{M^2} \right\|_F^2 \). By exploiting the linearity of the trace operator, the objective function can be expressed as
\begin{align}
    f &= \mathrm{tr}\left( \left(\frac{1}{\tau_k} \mathbf{\Theta}_{\mathrm{mc},k} \mathbf{\Theta}_{\mathrm{mc},k}^{\mathrm{H}} - \mathbf{I}_{M^2} \right)^{\mathrm{H}} \left( \frac{1}{\tau_k}\mathbf{\Theta}_{\mathrm{mc},k} \mathbf{\Theta}_{\mathrm{mc},k}^{\mathrm{H}} - \mathbf{I}_{M^2} \right) \right) \notag \\
    &= \mathrm{tr}\left( \left( \frac{1}{\tau_k} \mathbf{\Theta}_{\mathrm{mc},k} \mathbf{\Theta}_{\mathrm{mc},k}^{\mathrm{H}}\right)^2\right) -2\mathrm{tr}\left(\frac{1}{\tau_k}\mathbf{\Theta}_{\mathrm{mc},k} \mathbf{\Theta}_{\mathrm{mc},k}^{\mathrm{H}}\right) + M^2 .
\end{align}
The Euclidean gradient of the objective function \( f \) with respect to \( \mathbf{\Theta}_{\mathrm{mc},k} \) is given by
\begin{align}
    \frac{\partial f}{\partial \mathbf{\Theta}_{\mathrm{mc},k}} = \frac{4}{\tau_k^2} \mathbf{\Theta}_{\mathrm{mc},k} \left( \mathbf{\Theta}_{\mathrm{mc},k}^{\mathrm{H}} \mathbf{\Theta}_{\mathrm{mc},k} - \tau_k \mathbf{I}_{\tau_k} \right) .
    \label{eq43}
\end{align}
Accordingly, the partial derivative with respect to \(\boldsymbol{\eta}_t\) is \(\frac{\partial f}{\partial \boldsymbol{\eta}_t} = \frac{4}{\tau_k^2} \mathbf{\Theta}_{\mathrm{mc},k} \left[ \mathbf{\Theta}_{\mathrm{mc},k}^{\mathrm{H}} \mathbf{\Theta}_{\mathrm{mc},k} - \tau_k \mathbf{I}_{\tau_k} \right]_{:,t}\). To compute the element-wise derivatives in Eq.~\eqref{eq43}, the partial derivative of \( \mathbf{B}_t = \left( \mathbf{\Gamma}_t^{-1} - \mathbf{S} \right)^{-1} \) with respect to \( \left[ \boldsymbol{\gamma}_{t} \right]_{m}\) is evaluated first. Using the matrix inverse differentiation identity yields
\begin{subequations}
    \begin{align}
        \frac{\partial \mathbf{\Gamma}_t^{-1}}{\partial \left[ \boldsymbol{\gamma}_{t} \right]_{m}} &= -\mathbf{\Gamma}_t^{-1} \left( \frac{\partial \mathbf{\Gamma}_t}{\partial \left[ \boldsymbol{\gamma}_{t} \right]_{m}} \right) \mathbf{\Gamma}_t^{-1} = -\left[ \boldsymbol{\gamma}_{t} \right]_{m}^{-2} \mathbf{E}_m,
        \label{eq44a} \\
        \frac{\partial \mathbf{B}_t}{\partial \left[ \boldsymbol{\gamma}_{t} \right]_{m}} &= -\mathbf{B}_t \left( \frac{\partial \left( \mathbf{\Gamma}_t^{-1} - \mathbf{S} \right)}{\partial \left[ \boldsymbol{\gamma}_{t} \right]_{m}} \right) \mathbf{B}_t = \left[ \boldsymbol{\gamma}_{t} \right]_{m}^{-2} \mathbf{B}_t \mathbf{E}_m \mathbf{B}_t,
        \label{eq44b}
    \end{align}
\end{subequations}
where \( \mathbf{E}_m = \mathrm{Diag}(\mathbf{e}_m) \in \mathbb{R}^{M \times M} \), and 
\( \mathbf{e}_m \in \mathbb{R}^{M \times 1} \) is the standard basis vector whose \( m \)-th entry is 1 and all others are zero. By applying the vectorization identity, we derive
\begin{align}
    \frac{\partial \boldsymbol{\eta}_t}{\partial \left[ \boldsymbol{\gamma}_{t} \right]_{m}} = \left[ \boldsymbol{\gamma}_{t} \right]_{m}^{-2} (\mathbf{B}_t^{\mathrm{T}} \otimes \mathbf{B}_t) \mathrm{vec}(\mathbf{E}_m).
    \label{eq45}
\end{align}
Since \( \mathrm{vec}(\mathbf{E}_m) \in \mathbb{R}^{M^2 \times 1}\) contains only a single nonzero entry at position \( (m - 1)M + m \), we apply the chain rule and matrix calculus identities to compute the partial derivative of \( f \) with respect to \( \left[ \boldsymbol{\gamma}_{t} \right]_{m}^* \) as
\begin{align}
    \mathbf{G}_{m,t}^{\mathrm{E}} \triangleq\frac{\partial f}{\partial \left[ \boldsymbol{\gamma}_{t} \right]_{m}^*} = \frac{4}{\tau_k^2} \left[ \boldsymbol{\gamma}_{t} \right]_{m}^{-2} \left[ \mathbf{B}_t^{*} \odot \mathbf{R}_t \right]_{m,m},
    \label{eq46}
\end{align}
where \( \mathbf{R}_t = \mathrm{mat} \left( \mathbf{\Theta}_{\mathrm{mc},k} \left[ \mathbf{\Theta}_{\mathrm{mc},k}^{\mathrm{H}} \mathbf{\Theta}_{\mathrm{mc},k} - \tau_k\mathbf{I}_{\tau_k} \right]_{:,t} \right) \in \mathbb{C}^{M \times M} \), and \( \mathbf{G}^{\mathrm{E}} \in \mathbb{C}^{M \times \tau_1} \) denotes the Euclidean gradient matrix. To enforce the unit-modulus constraint \( \left[ \boldsymbol{\gamma}_{t} \right]_{m} \in \mathbb{S}^1_{\mathbb{C}} \), the Euclidean gradient is projected onto the tangent space of the complex circle manifold, yielding the Riemannian gradient matrix \(\mathbf{G}^{\mathrm{R}} \in \mathbb{C}^{M \times \tau_k}\) as follows
\begin{align}
    \mathbf{G}_{:,t}^{\mathrm{R}} = \mathbf{G}_{:,t}^{\mathrm{E}} - \mathrm{Re} \left\{ \mathbf{G}_{:,t}^{\mathrm{E}} \odot \boldsymbol{\gamma}_t^* \right\} \odot \boldsymbol{\gamma}_t.
\end{align}
The resulting optimization problem is solved using the Polak–Ribi\`ere conjugate gradient method on the Riemannian manifold. The initial value of \(\mathbf{\Theta}_{\mathrm{cv},k} \) (\( \forall k \in \mathcal{K} \)) is set to the random Bernoulli matrix, with each element independently drawn from \(\{-1, +1\}\) with equal probability~\cite{ref16}.

\section{Simulation Results}\label{simulation}

In this section, we present simulation results to evaluate the performance of the proposed MC-aware three-stage channel estimation protocol for the RIS-aided MU-MISO system. 
\begin{itemize}
    \item \textit{System Configuration:} The RIS is centered at \(\left(0,0,0\right)\) and faces the positive \(x\)-axis. The BS is located at \(100~\mathrm{m} \times \left( \sin \frac{5\pi}{6}, \cos \frac{5\pi}{6}, 0 \right)\). The users are centered at \(10~\mathrm{m} \times \left( \sin \frac{\pi}{3}, \cos \frac{\pi}{3}, 0 \right)\), with the number of users set to \(K=4\). The complex channel gains are modeled as \(\alpha_{l} \sim \mathcal{CN}(0,10^{-3} d_{\mathrm{BR}}^{-2.2})\) and \(\beta_{k,j} \sim \mathcal{CN}(0,10^{-3} d_{\mathrm{RU}}^{-2.8})\), where \(d_{\mathrm{BR}}\) and \(d_{\mathrm{RU}}\) denote the BS-RIS and RIS-user distances, respectively. The BS and RIS employ UPAs of size \(N_h \times N_v = 8 \times 8\) and \(M_h \times M_v = 4 \times 4\), respectively. The number of propagation paths is set to \(L=4\) for the BS-RIS link and \(J_1=\cdots=J_K=J=2\) for all RIS-user links. The antenna and inter-element spacing at the BS and the RIS are both set to \(d_{\mathrm{BS}} = d_{\mathrm{RIS}} = \lambda_{\mathrm{c}}/2\).    
    \item \textit{Mutual Coupling Model:} The carrier frequency is set to \(f_c = 28~\mathrm{GHz}\). The RIS passive elements are modeled as cylindrical thin wires of perfectly conducting material, each having a length \(l = \lambda_{\mathrm{c}}/32\) and a radius \(a = \lambda_{\mathrm{c}}/500\), where \(a \ll l\) to satisfy the thin-wire approximation. The characteristic impedance is assumed to be \(Z_0 = 50 \Omega\).  
    \item \textit{Noise and Power Setting:} The thermal noise power at both the BS and the RIS is fixed at \(\sigma_1^2 = \sigma_2^2 = -80~\mathrm{dBm}\). The transmit power is set to \(p = 25\text{dBm}\).
\end{itemize}
To evaluate the performance of the proposed channel estimation protocol, we adopt the normalized mean square error (NMSE) as the primary metric, defined as
\[
\text{NMSE} = \mathbb{E} \left\{ \frac{\sum_{k=1}^{K} \| \widehat{\mathbf{Y}}_{i,k} - \bar{\mathbf{Y}}_{k} \|_F^2}{\sum_{k=1}^{K} \| \bar{\mathbf{Y}}_{k} \|_F^2} \right\},
\]
where \( \bar{\mathbf{Y}}_{k} \) denotes the noise-free received signal for user \(k\), and the subscript \(i \in \{\mathrm{mc}, \mathrm{cv}\}\) indicates whether the MC-aware or conventional channel model is employed. Specifically, the reconstructed signals are given by \(\widehat{\mathbf{Y}}_{\mathrm{mc},k} = \sqrt{p} \widehat{\mathbf{G}}_{\mathrm{mc},k} \mathbf{\Theta}_{\mathrm{mc},k}\) and \(\widehat{\mathbf{Y}}_{\mathrm{cv},k} = \sqrt{p} \widehat{\mathbf{G}}_{\mathrm{cv},k} \mathbf{\Theta}_{\mathrm{cv},k}\).

A comprehensive performance evaluation is conducted to compare several algorithms integrated into the proposed MC-aware three-stage channel estimation protocol. The following summarizes the key features of each algorithm.

\begin{itemize}
    \item \textbf{Proposed~1} : Proposed~1 adopts the Root-MUSIC algorithm as the dimension-reduced subspace-based common AoA estimation, as implemented in Algorithm~\ref{alg1} during Stage~I. Root-MUSIC identifies the roots of a polynomial that lie closest to the unit circle in the complex plane, where the angular positions of these roots directly correspond to the DOAs~\cite{ref21}.
    \item \textbf{Proposed~2} : Proposed~2 employs the TLS-ESPRIT algorithm as the dimension-reduced subspace-based common AoA estimation, also implemented in Algorithm~\ref{alg1} during Stage~I. TLS-ESPRIT refines the solution to the rotational invariance equation using the TLS criterion, which accounts for perturbations in both the signal subspace and the observed data~\cite{ref22}.
    \item \textbf{MC-unaware} : The MC-unaware algorithm employs the Root-MUSIC algorithm for common AoA estimation and uses the estimation strategy described in Section~\ref{unaware} for estimating the remaining component \(\mathbf{H}_{\mathrm{RIS}}^{\mathrm{cv},k}\).
    \item \textbf{Direct-OMP} : Direct-OMP formulates the vectorized received signal as \(\mathrm{vec} \left(\mathbf{Y}_{k} \right)= \sqrt{p} \left(\mathbf{\Theta}_{k}^{\mathrm{T}} \otimes \mathbf{I}_{N} \right) \mathrm{vec} \left(\mathbf{G}_{k}\right) + \mathrm{vec} \left(\mathbf{N}_{k}\right)\), based on direct vectorization of Eq.~\eqref{eq12}~\cite{ref18}. The estimation problem is then solved via \(L^2J_k\)-sparse signal recovery problem, which incurs both high computational complexity and significant pilot overhead\footnote{Specifically, \(\mathrm{vec} \left(\mathbf{G}_{k}\right) = \left( \left(\mathbf{A}_{M,k}^{\mathrm{T}} \otimes \mathbf{A}_{M}^{\mathrm{H}}\right)^{\mathrm{T}} \otimes \mathbf{A}_{N}\right) \mathrm{vec}\left( \boldsymbol{\beta}_{k}^{\mathrm{T}} \otimes \mathbf{\Lambda} \right)\). By optimizing the RIS phase shift training matrix as described in Section~\ref{RIS}, the orthogonality of the equivalent dictionary in this formulation is also improved, enabling a more equitable performance comparison.}.
    \item \textbf{SBL} : Sparse Bayesian learning can be employed to solve the sparse signal recovery problem. The equivalent measurement matrix \(\breve{\mathbf{Y}}_{k}\) in Eq.~\eqref{eq19} is recovered using SBL\footnote{The SBL framework adopts an expectation–maximization (EM) procedure for sparse vector estimation. Specifically, in the \(j\)-th iteration, the E-step computes \(\widehat{\mathbf{B}}_{k} = \mathbf{\Sigma}\mathbf{D}_{k}^{\mathrm{H}}\mathbf{R}^{-1}\breve{\mathbf{Y}}_{k}\) with the error covariance matrix \(\mathbf{\Sigma} = (\mathbf{D}_{k}^{\mathrm{H}}\mathbf{R}^{-1}\mathbf{D}_{k} + \mathbf{\Xi}^{-1})^{-1}\), where \(\mathbf{R}\) is the corresponding noise covariance matrix. While the M-step updates the hyperparameters as \(\widehat{\xi}^{(j)}_g = \boldsymbol{\Sigma}^{(j)}(g,g) + \frac{1}{L} \sum_{l=1}^{L} \Big| \widehat{\mathbf{B}}_{k}^{(j)}(g,l)\Big|^2\), where \(\mathbf{\Xi} = \mathrm{Diag} \left( \xi_1, \cdots, \xi_{DD_{k}} \right)\).}, after eliminating the common AoA effect via the Root-MUSIC algorithm. The algorithm is executed with the maximum number of iterations set to 50 and the convergence threshold set to \(10^{-6}\).
\end{itemize}

\begin{figure}
    \centering
    \includegraphics[width=3.5in]{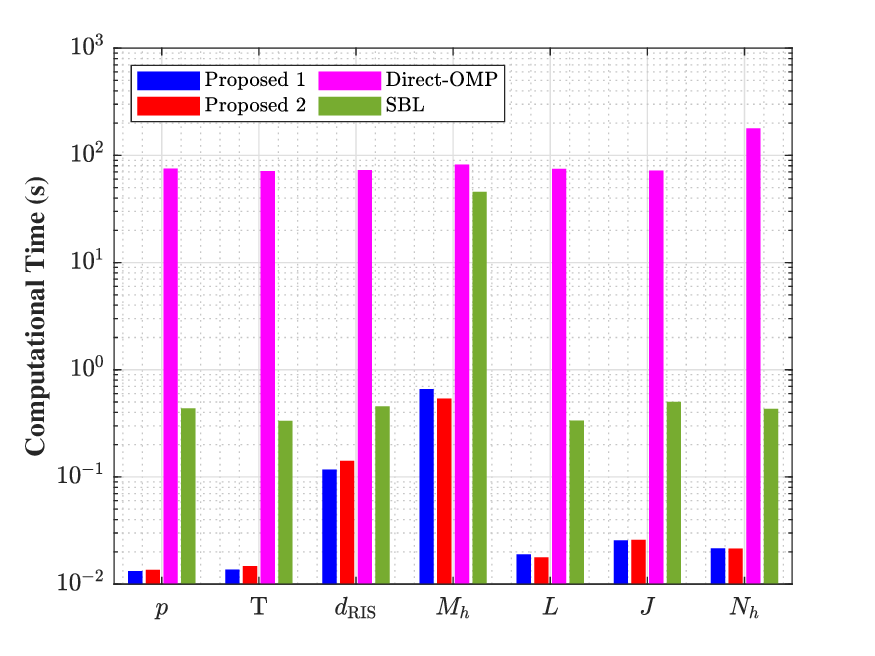}
    \caption{Computational time of the algorithms under different parameter settings.}
\label{time}
\end{figure}

To facilitate a direct comparison of computational complexity, Fig.~\ref{time} presents the computational time per channel simulation under varying parameters. As illustrated in the figure, several system parameters have a greater impact the computational cost. Specifically, the RIS inter-element spacing \(d_{\mathrm{RIS}}\) and the horizontal RIS dimension \(M_h\) affect the scattering matrix \(\mathbf{S}\) that characterizes the MC effect. In addition, \(M_h\) directly determines the size of the equivalent dictionary matrix. Moreover, an increased number of propagation paths \(L\) or \(J\) leads to higher per-simulation runtime and will affect simulation performance, as further demonstrated in subsequent analysis. As shown in the figure, the proposed methods exhibit substantially lower computational complexity compared to the other algorithms.

\begin{figure}
    \centering
    \includegraphics[width=3.5in]{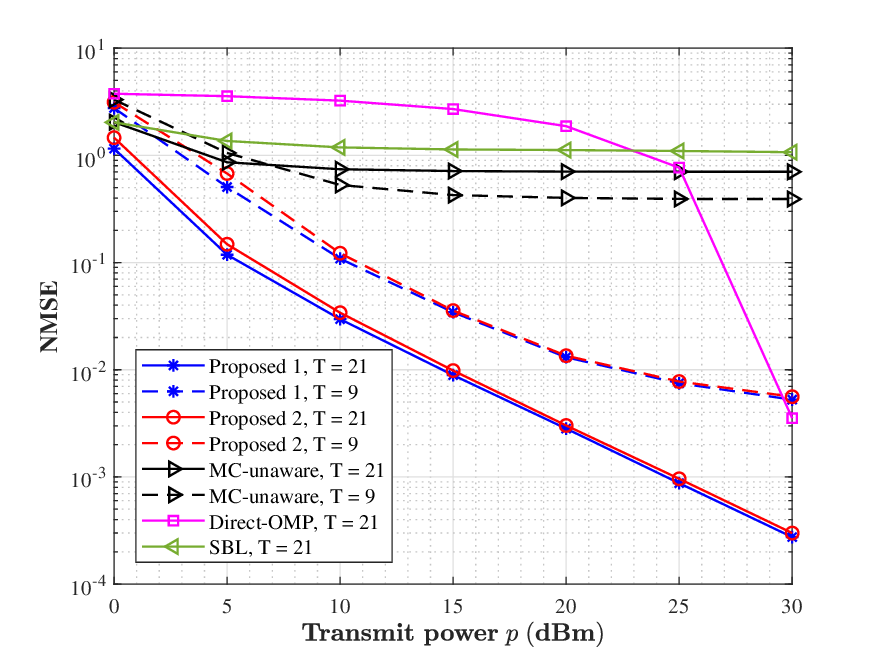}
    \caption{NMSEs vs. Transmit power \(p\).}
\label{fig1}
\end{figure}

Fig.~\ref{fig1} illustrates the NMSE performance as a function of the transmit power \(p\), evaluated under two average pilot overhead settings: \(\mathrm{T} = 21\) and \(\mathrm{T} = 9\). In the case of \(\mathrm{T} = 21\), the typical user (user~1) is allocated \(\tau_1 = 24\) pilots, while each other user is assigned \(\tau_k = 20\); for \(\mathrm{T} = 9\), \(\tau_1 = 12\) and \(\tau_k = 8\). The larger allocation to the typical user is intended to mitigate error propagation in the three-stage estimation process. The results show that both Proposed~1 and Proposed~2 achieve a substantial NMSE reduction as \(p\) increases. At lower transmit powers, Proposed~1 slightly outperforms Proposed~2. In contrast, the Direct-OMP algorithm suffers from high NMSE at low power due to its high-dimensional sparse recovery formulation. It only reaches comparable performance to the proposed methods when the transmit power reaches \(30\) dBm. Therefore, to ensure fair comparisons in subsequent simulations, the transmit power is fixed at \(p = 25\) dBm. The SBL method maintains high NMSE across the transmit power range. Its performance is hindered by the difficulty of accurately estimating noise covariance in RIS-aided system model. Meanwhile, the MC-unaware methods consistently underperform, further highlighting the importance of modeling MC-awrare in RIS-aided models.

\begin{figure}
    \centering
    \includegraphics[width=3.5in]{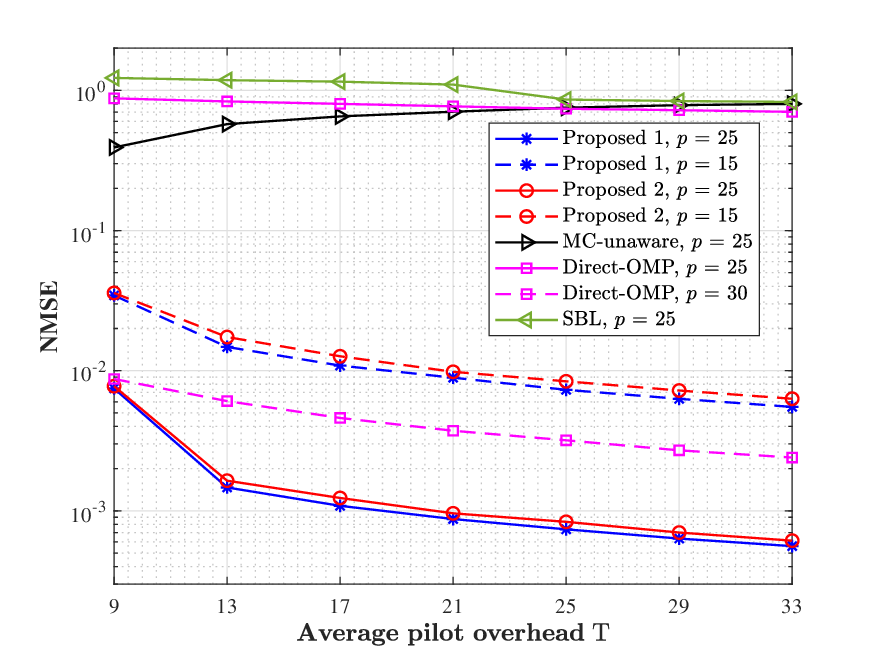}
    \caption{NMSEs vs. Average pilot overhead \(\mathrm{T}\).}
\label{fig2}
\end{figure}

Fig.~\ref{fig2} presents the NMSE performance as a function of the average pilot overhead \(\mathrm{T}\), with two transmit power levels \(p = 25\) dBm and \(p = 15\) dBm for the proposed methods. The figure demonstrates that the NMSEs of Proposed~1 and Proposed~2 decrease monotonically as the pilot overhead increases, indicating improved estimation accuracy with more pilot resources. Notably, Proposed~1 achieves slightly better performance than Proposed~2. Both Direct-OMP and SBL exhibit gradual improvements as \(\mathrm{T}\) increases. However, their estimation accuracy remains significantly inferior to that of the proposed approaches, even when the proposed methods operate at the lower transmit power (\(p = 15\) dBm), underscoring their higher pilot and power requirements. Direct-OMP’s poor performance stems from its reliance on full vectorization and large-scale sparse recovery, which imposes stringent demands on pilot overhead. Interestingly, the MC-unaware method's NMSEs worsen as pilot overhead grows. This phenomenon, also observed in Fig.~\ref{fig1}, is attributed to the model mismatch that becomes increasingly pronounced as pilot dimensions expand. Specifically, the divergence between the conventional RIS channel model and the MC-aware formulation grows with the dimensionality of the RIS phase shift training matrices \(\mathbf{\Theta}_{\mathrm{cv},k}\) and \(\mathbf{\Theta}_{\mathrm{mc},k}\), leading to increased estimation errors.

\begin{figure}
    \centering
    \includegraphics[width=3.5in]{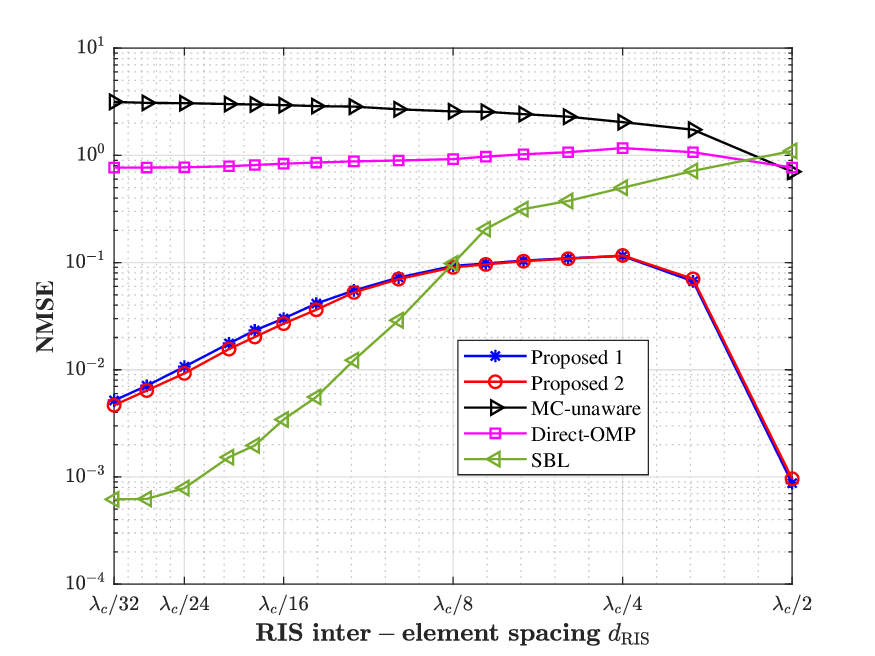}
    \caption{NMSEs vs. RIS inter-element spacing \(d_{\mathrm{RIS}}\) with \(p\) = 25 dBm.}
\label{fig3}
\end{figure}

Fig.~\ref{fig3} illustrates the NMSE performance as a function of the RIS inter-element spacing \(d_{\mathrm{RIS}}\), which controls the severity of MC effects. As the inter-element spacing decreases, the proposed MC-aware methods exhibit a non-monotonic NMSE trend with decreasing RIS inter-element spacing, initially increasing before declining. This turning behavior can be attributed to two competing factors: (i) the impact of \(d_{\mathrm{RIS}}\) on the orthogonality of the equivalent dictionary matrix under MC, and (ii) the direct influence of \(d_{\mathrm{RIS}}\) on the array response characteristics that affect sparse recovery performance. Specifically, when \(d_{\mathrm{RIS}} \le \lambda_c/5\), increasing \(d_{\mathrm{RIS}}\) aggravates the MC-induced loss of dictionary orthogonality, thereby worsening sparse recovery and increasing the NMSE. When \(d_{\mathrm{RIS}} \ge \lambda_c/5\), the degradation in dictionary orthogonality tends to saturate, whereas the improvement in array response characteristics with increasing spacing becomes dominant, leading to a decreasing NMSE. In particular, NMSE peaks around \(d_{\mathrm{RIS}} \in [\lambda_c/5,\, \lambda_c/3]\). Nevertheless, both Proposed~1 and Proposed~2 consistently outperform Direct-OMP across all spacing levels. Notably, SBL performs relatively well under small RIS inter-element spacing conditions. The MC-unaware method displays a steadily degrading trend as inter-element spacing decreases, underscoring its vulnerability to model mismatch under increasing MC severity. This highlights the importance of incorporating MC-aware modeling, particularly when dense RIS deployments are considered.

\begin{figure}
    \centering
    \includegraphics[width=3.5in]{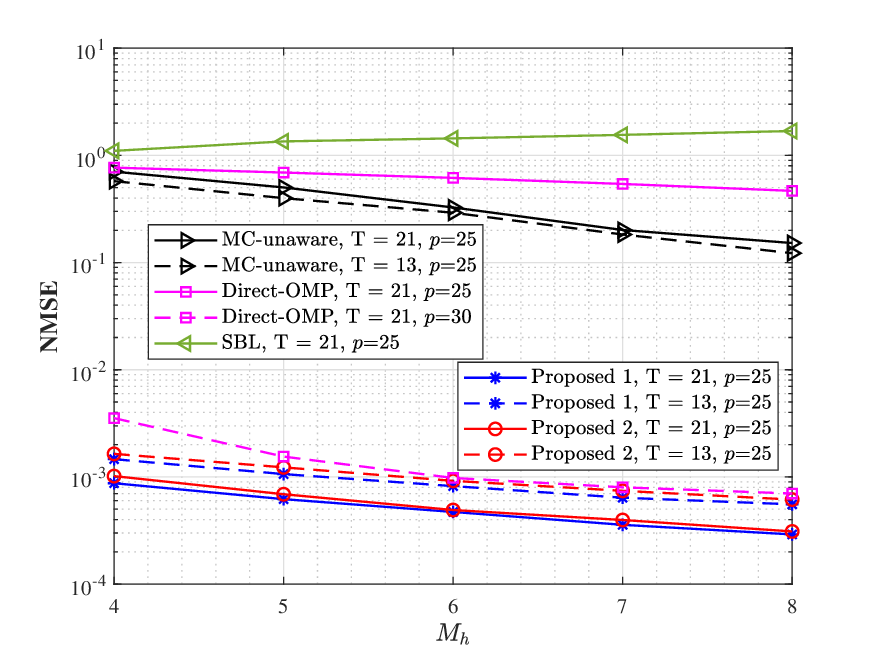}
    \caption{NMSEs vs. Number of elements at the RIS \(M = M_h \times M_v\) with \(p\) = 25 dBm.}
\label{fig4}
\end{figure}

Fig.~\ref{fig4} presents the NMSE performance as a function of the number of RIS elements along the horizontal dimension \(M_h\), under two average pilot overhead settings: \(\mathrm{T} = 21\) and \(\mathrm{T} = 13\). Increasing \(M_h\) expands the RIS aperture while maintaining fixed inter-element spacing, thereby reducing the MC effect. This structural change explains the observed trend: the NMSEs of proposed methods gradually decrease with increasing \(M_h\). In low \(M_h\) regimes Proposed~2 slightly outperforms Proposed~1. The Direct-OMP and MC-unaware methods continue to exhibit high NMSE values across all settings. Notably, the performance of SBL worsens as \(M_h\) grows. This degradation is attributed to the increasing dimensionality of the equivalent dictionary matrix \( \mathbf{D}_k \), which complicates convergence and elevates estimation error in iteration.

\begin{figure}
    \centering
    \includegraphics[width=3.5in]{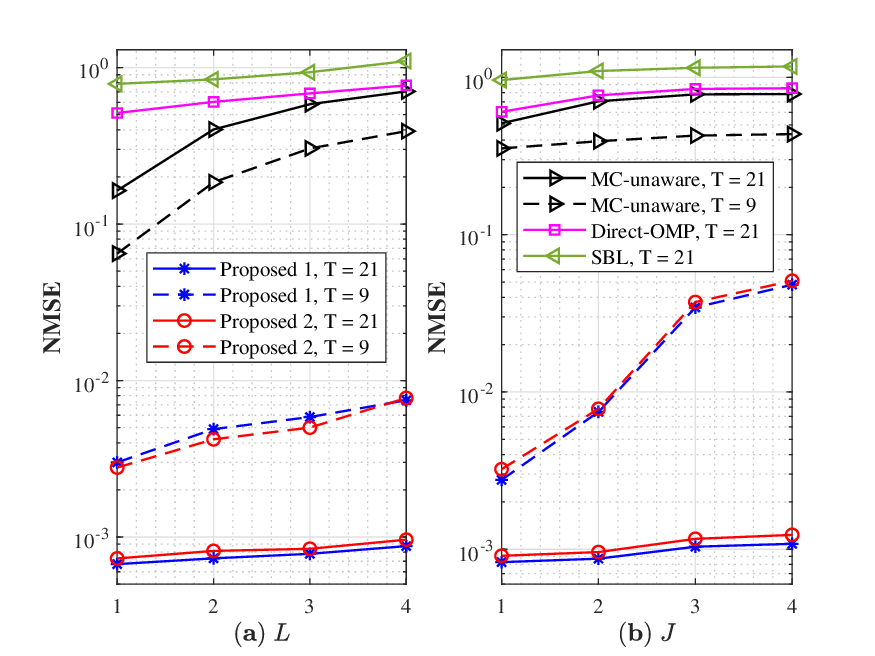}
    \caption{NMSEs vs. Numbers of propagation paths in the RIS-BS link \(L\)~(a) and user-RIS links \(J\)~(b) with\(p\) = 25 dBm.}
\label{fig5}
\end{figure}

Fig.~\ref{fig5}~(a) depicts the impact of the numbers of propagation paths in the RIS-BS link \(L\) on the NMSE performance, under two average pilot overhead settings: \(\mathrm{T} = 21\) and \(\mathrm{T} = 9\). As \(L\) increases, all methods exhibit consistent performance degradation. This degradation is primarily attributed to two factors: (i) the growth in the number of unknown channel parameters (e.g., AoAs at the BS, AoDs at the RIS, and path gains), which exacerbates the estimation difficulty under fixed pilot overhead; and (ii) the additional uncertainty associated with estimating the number of paths itself. Despite the performance drop at larger \(L\), the proposed methods consistently outperform the other methods across all settings. Fig.~\ref{fig5}~(b) illustrates the impact of the numbers of propagation paths in the user-RIS links \(J\) on the NMSE performance. As \(J\) increases, the NMSE rises accordingly, reflecting greater estimation difficulty, consistent with the explanation in Fig.~\ref{fig5}~(a). The degradation is more pronounced than that caused by increasing \(L\), since \(J\) directly determines the sparsity level in the sparse recovery problems.

\begin{figure}
        \centering
        \includegraphics[width=3.5in]{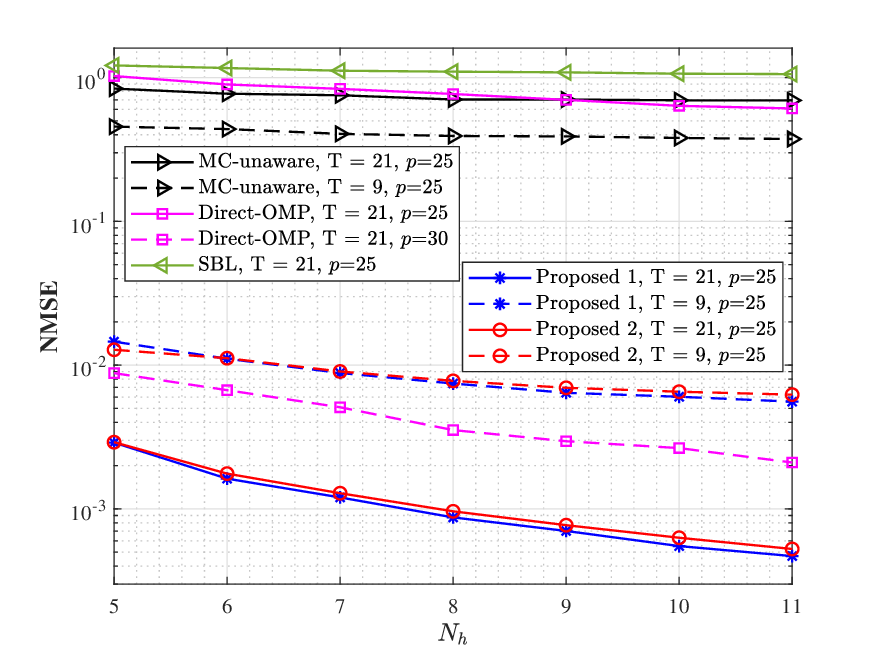}
        \caption{NMSEs vs. Number of antennas at the BS \(N = N_h \times N_v\) with \(p\) = 25 dBm.}
    \label{fig6}
\end{figure}

Fig.~\ref{fig6} depicts the NMSE performance versus the number of BS antennas along the horizontal dimension \(N_h\) under two average pilot overhead settings, namely \(\mathrm{T} = 21\) and \(\mathrm{T} = 9\). It is observed that the NMSE of all considered methods decreases as \(N_h\) increases. This behavior is primarily attributed to the improved accuracy of the dimension-reduced subspace-based common AoA estimation, in which \(N_h\) plays a critical role. Moreover, the proposed methods consistently outperform the benchmark approaches, even under low pilot overhead conditions, with the performance gap becoming more pronounced as \(N_h\) increases. These results further indicate that the proposed framework more effectively leverages the enhanced common AoA estimation capability to reduce the overall channel estimation error.

\begin{figure}
    \centering
    \includegraphics[width=3.5in]{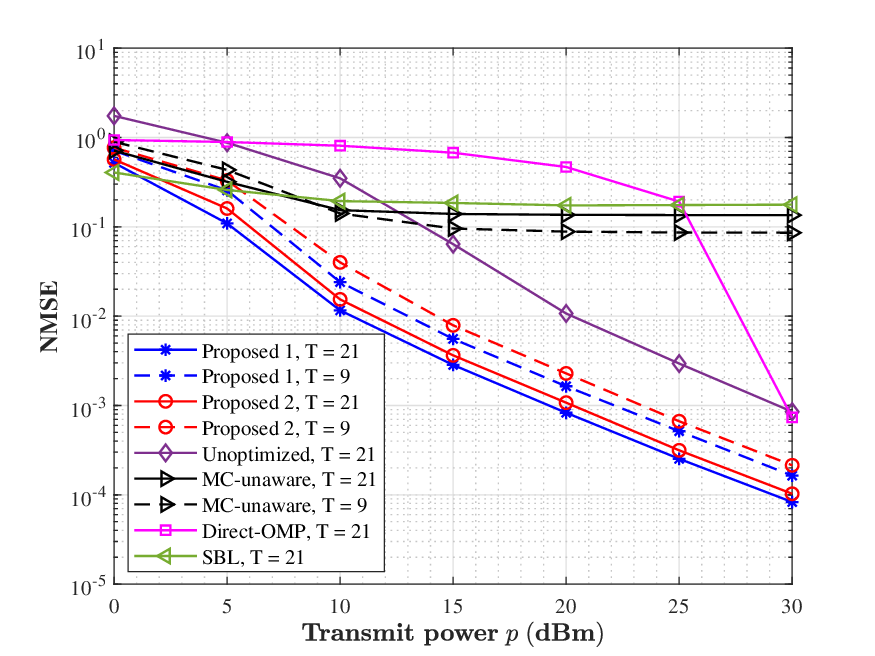}
    \caption{NMSEs vs. Transmit power \(p\) (\(K=1\)).}
\label{fig7}
\end{figure}

Fig.~\ref{fig7} examines the impact of optimizing the RIS phase shift training matrix on estimation performance. Specifically, it shows the NMSE performance as a function of transmit power \(p\) in a single-user scenario (\(K=1\)), evaluated under two pilot overhead settings, \(\mathrm{T} = 21\) and \(\mathrm{T} = 9\). The Unoptimized scheme employs the Root-MUSIC algorithm in Stage~I but does not employ optimized RIS phase shift training matrix; instead, it relies on the random phase shift matrix. The performance gap between the Unoptimized scheme and Proposed~1 demonstrates the performance gains achieved through RIS phase shift training matrix optimization. Moreover, the trends of the other algorithms with respect to transmit power are consistent with those observed in Fig.~\ref{fig1}.

\section{Conclusion}\label{conclusion}

In this paper, we proposed an MC-aware three-stage channel estimation protocol for RIS-aided MU-MISO mmWave systems, which accurately captures the cascaded channel characteristics under practical RIS deployment conditions. We leveraged a dimension-reduced subspace-based method to achieve accurate estimation of common AoA at the BS in Stage~I. Then an OMP-based approach was developed to enable low-complexity estimation of the cascaded channel for the typical user, while simultaneously estimating common AoD at the RIS. We further reduced pilot overhead for other users by leveraging the fact that all users share the common RIS–BS channel. Additionally, we proposed a Riemannian manifold optimization framework to design the RIS phase shift training matrix by explicitly accounting for the MC effect, achieving clear performance gains over random phase shift scheme. Simulation results validate that proposed method significantly outperforms MC-unaware and existing approaches in terms of both estimation accuracy and pilot efficiency.

\bibliographystyle{IEEEtran}
\bibliography{ref}
\balance



\end{document}